\documentclass[fleqn,usenatbib]{mnras}

\usepackage{newtxtext,newtxmath}
\usepackage[T1]{fontenc}

\DeclareRobustCommand{\VAN}[3]{#2}
\let\VANthebibliography\thebibliography
\def\thebibliography{\DeclareRobustCommand{\VAN}[3]{##3}\VANthebibliography}

\usepackage{graphicx}	% Including figure files
\usepackage{amsmath}	% Advanced maths commands
\usepackage{gensymb}   % Degree Symbol
\usepackage{siunitx}
\usepackage{threeparttable}
\newcommand{\Autoref}[1]{%
  \begingroup%
  \def\chapterautorefname{Chapter}%
  \def\sectionautorefname{Section}%
  \def\subsectionautorefname{Subsection}%
  \autoref{#1}%
  \endgroup%
}
\title[Cold extragalactic gas clouds at $z=0.45$]{Observations of cold extragalactic gas clouds at $z = 0.45$ towards PKS 1610-771}

\author[Simon Weng et al.]{
Simon Weng,$^{1,2,3}$\thanks{E-mail: swen2649@uni.sydney.edu.au (KTS)}
Elaine M. Sadler,$^{1,2,3}$
Caroline Foster,$^{1,3}$
C\'eline P\'eroux,$^{4,5}$ 
\newauthor
Elizabeth K. Mahony,$^{2}$
James R. Allison,$^{3,6}$
Vanessa A. Moss,$^{2,1}$
Renzhi Su,$^{7,8,2}$ 
\newauthor
Matthew T. Whiting$^{2}$
and Hyein Yoon$^{1,3}$
\\
$^1$Sydney Institute for Astronomy, School of Physics, University of Sydney, NSW 2006, Australia\\
$^2$CSIRO Space and Astronomy, Australia Telescope National Facility, PO Box 76, Epping, NSW 1710, Australia \\
$^3$ARC Centre of Excellence for All Sky Astrophysics in 3 Dimensions (ASTRO 3D) \\
$^4$European Southern Observatory, Karl-Schwarzschildstrasse 2, D-85748 Garching bei M\"{u}nchen, Germany \\
$^5$Aix Marseille Universit\'{e}, CNRS, LAM (Laboratoire d’Astrophysique de Marseille) UMR 7326, F-13388 Marseille, France \\
$^6$Sub-Dept. of Astrophysics, Department of Physics, University of Oxford, Denys Wilkinson Building, Keble Rd., Oxford, OX1 3RH, UK\\
$^7$ Key Laboratory for Research in Galaxies and Cosmology, Shanghai Astronomical Observatory, Chinese Academy of Sciences, 80 Nandan Road, \\ Shanghai 200030, China \\
$^8$ University of Chinese Academy of Sciences, 19A Yuquan Road, Beijing 100049, China \\
}
\date{Accepted XXX. Received YYY; in original form ZZZ}

\pubyear{2021}

\begin{document}
\label{firstpage}
\pagerange{\pageref{firstpage}--\pageref{lastpage}}
\maketitle

\begin{abstract}
We present results from MUSE observations of a 21-cm \ion{H}{i} absorption system detected with the Australian Square Kilometre Array Pathfinder radio telescope at redshift $z = 0.4503$ towards the $z=1.71$\ quasar PKS 1610-771. 
We identify four galaxies (A, B, X and Y) at the same redshift as the $21$-cm \ion{H}{i} Damped Lyman-$\alpha$ (DLA) absorption system, with impact parameters ranging from less than 10\,kpc to almost 200\,kpc from the quasar sightline.
\ion{Ca}{ii} and \ion{Na}{i} absorption is seen in the MUSE spectrum of the background QSO, with velocities coinciding with the initial \ion{H}{i} $21$-cm detection, but tracing less dense and warmer gas. 
This metal-line component aligns with the rotating ionised disc of galaxy B (impact parameter $18$ kpc from the QSO) and appears to be co-rotating with the galaxy disc.
In contrast, the $21$-cm \ion{H}{i} absorber is blueshifted relative to the  galaxies nearest the absorber and has the opposite sign to the velocity field of galaxy B.
Since galaxies A and B are separated by only $17$ kpc on the sky and $70$ km s$^{-1}$ in velocity, it appears likely that the $21$-cm detection traces extragalactic clouds of gas formed from their interaction.
This system reveals that the cold $100$ K neutral gas critical for star formation can be associated with complex structures beyond the galaxy disc, and is a first case study made in preparation for future large $21$-cm absorption surveys like the ASKAP First Large Absorption Survey in \ion{H}{i}.
\end{abstract}

\begin{keywords}
intergalactic medium -- galaxies: kinematics and dynamics -- quasars: absorption lines -- radio lines: ISM
\end{keywords}

%%%%%%%%%%%%%%%%%%%%%%%%%%%%%%%%%%%%%%%%%%%%%%%%%%

%%%%%%%%%%%%%%%%% BODY OF PAPER %%%%%%%%%%%%%%%%%%

\section{Introduction}
The reservoirs of cold gas within galaxies serve as fuel for star formation, which in turn drives the morphological and kinematic properties of galaxies.
It follows that the availability of hydrogen and phenomena that affect this availability play an important role in how galaxies evolve \citep{PerouxHowk2020}.
Placing observational constraints on the amount and distribution of gas across all redshifts will improve our understanding of galaxy evolution.
Processes that affect gas abundance such as inflows and outflows also require characterisation from observations as we pursue more complete simulations of the baryon cycle.

The amount and distribution of neutral hydrogen (\ion{H}{i}) is well-constrained at low redshift ($z \lesssim 0.1$) from \ion{H}{I} emission-line surveys \cite{Rosenberg2002, Zwaan2005, Hoppmann2015, GiovanelliHaynes2015, Jones2018}.
Beyond this redshift, $21$-cm emission is too faint for detection without stacking techniques \citep[e.g.][]{Kanekar2016, Rhee2018, Chowdhury2020} as the transition rate is highly forbidden.
While \ion{H}{i} emission-line stacking can extend the search for neutral gas beyond the local Universe, the amount and distribution of gas still remains uncertain at intermediate redshifts ($0.2 \leq z \leq 1.7$).
At $z \gtrsim 1.7$, optical observations of Lyman-$\alpha$ $1215$ \AA \  absorption in background quasar spectra constrain the cosmic density of hydrogen \citep[e.g.][]{Noterdaeme2012}.
These damped Lyman-$\alpha$ absorbers (DLAs) trace column densities above $2 \times 10^{20} \ \text{atoms cm}^{-2}$ and account for $\sim 80\%$ of the cosmic neutral gas density \citep{Peroux2003, Noterdaeme2009, Noterdaeme2012, Zafar2013, Berg2019}.
This technique has been extended to lower redshifts by first selecting candidate systems with strong \ion{Mg}{ii} absorption and then following up with Hubble Space Telescope (HST) Ultraviolet spectroscopy to search for Lyman-$\alpha$ \citep{Rao2000, Rao2006, Rao2017}.
In such studies, the selection biases are difficult to quantify and the limited sample size results in the cosmic \ion{H}{i} mass density remaining poorly constrained.
Alternatively, untargeted surveys for DLAs in the UV spectra of quasars avoid potential biases \citep{Neeleman2016}, but are statistically limited by the small number of detections.

Analogous to Lyman-$\alpha$ studies in the spectra of quasars, the $21$-cm \ion{H}{i} absorption line is a useful tool for tracing neutral gas in the distant Universe. 
While methodologically similar (i.e. using a background source to probe sightlines through gas), $21$-cm absorption preferentially traces the coldest ($T_s \sim 100$ K) \ion{H}{i} and assumptions in the spin temperature are required to obtain a total column density.
Unlike \ion{H}{i} emission studies, absorption line spectroscopy is not limited by redshift, but rather by the intensity of the background source.
However, untargeted surveys of $21$-cm absorption have thus far been limited by poor spectral bandpass and terrestrial radio-frequency interference \citep{Brown1983, Darling2004}.
It is only with the recent construction of SKA (Square Kilometre Array) pathfinder telescopes Australian Square Kilometre Array Pathfinder (ASKAP) \citep{Johnston2007, DeBoer2009} and Meer Karoo Array Telescope (MeerKAT) \citep{JonasMeerKat2016}, with wide-band correlators and locations in radio-quiet sites, that these limitations are overcome. 

The First Large Absorption Survey in \ion{H}{i} \citep[FLASH,][]{allison2021} is an all-sky survey using the ASKAP radio telescope to search for neutral gas systems.
ASKAP comprises $36$ $12$-m antennas equipped with novel phase array feed technology and possesses a $30\ \text{deg}^2$ field-of-view \citep{Hotan2021}.
The FLASH project will search for \ion{H}{i} absorption in the largely unexplored redshift range $0.4 < z < 1.0$ and commissioning and early science results have already yielded several new detections \citep{Sadler2020, Allison2020}.
Such untargeted surveys for neutral hydrogen are unaffected by dust obscuration \citep{PontzenPettini2009, Krogager2019} and 
remain unbiased by spectroscopic pre-selection of targets using the \ion{Mg}{ii} absorption line \citep{Neeleman2016}.

While $21$-cm absorption line spectroscopy is a powerful technique for probing the neutral gas content of galaxies, it provides only a single sightline through systems.
The origin of the gas detected in intervening systems remains unclear without follow-up imaging of the galaxy or galaxies associated with the neutral gas.
Integral-field spectroscopy (IFS) provides both spectra and imaging for objects in a field and so allows efficient identification of galaxies associated with the absorbing gas.
The technique of combining UV absorption spectroscopy with IFS observations in the near-infrared has been applied at redshifts $z \sim 1$ and $z \sim 2$ with the SINFONI instrument \citep{Bouche2007, Peroux2011, Peroux2013}.
This allows properties such as the SFR, gas kinematics and metallicity of these systems to be analysed and used to interpret the physical behaviour of the gas seen initially in absorption (i.e. outflows, inflows or co-rotation with halo). 
More recently, the MusE GAs FLOw and Wind (MEGAFLOW) survey compared the properties of the gas traced by \ion{Mg}{ii} absorption with the kinematics, stellar masses, star formation rates (SFRs) and orientation of their associated galaxies at $z \approx 1$ to investigate gas accretion and galactic winds \citep{Schroetter2019, Zabl2019}.
The MUSE-ALMA Halos Survey study the environments of strong \ion{H}{i} absorbers at $z \leq 1.4$ and reveal an increasing number of galaxy groups associated with a single absorber \citep{Peroux2016, Peroux2019, Klitsch2017, Rahmani2018, Hamanowicz2020, Szakacs2021}.
These studies, using the complementary methods of absorption and 3D spectroscopy, reveal that the environment of absorbers is often complex and larger, more representative samples are necessary to understand the cycling of gas in these systems.

Thus far, absorbers at $0.2 \leq z \leq 1.7$ have column densities measured using HST observations of the Lyman-$\alpha$ line in the UV spectra of quasi-stellar objects (QSOs), which results in a limited sample for follow-up imaging and spectroscopy 
\citep{Neeleman2016, Rao2017}. 
We expect to detect $\sim 1000$ $21$-cm absorbers with the FLASH project, providing a much larger sample of absorbers that will allow us to examine the relationship between the galaxy environment and neutral gas behaviour at intermediate redshift.
Moreover, in contrast to previous studies that use the \ion{Mg}{ii} doublet in absorption to trace gas that is $T \sim 1000$ K in the circumgalactic medium (CGM), \ion{H}{i} $21$-cm absorption traces the cold neutral gas that is required for star formation.
Mapping the coldest neutral gas in the CGM of galaxies enables us to directly explore the effects of accretion, outflows and interactions on star formation in galaxies.
Here, we present new MUSE observations of galaxies associated with a $21-$cm DLA at $z_{\rm abs}=0.4503$ detected from a ASKAP commissioning project \citep{Sadler2020} towards the $z=1.71$ quasar PKS 1610-771 \citep{Hunstead1980}.
\Autoref{sec: data} presents the initial \ion{H}{i} detection and relevant ancillary observations.
Details of the data processing and additional sky corrections for the new MUSE observations are found in \autoref{sec: MUSE_data}.
In \autoref{sec: analysis}, we analyse the MUSE observations and identify objects in the field associated with the absorber.
Finally, we discuss the physical nature of the gas detected with ASKAP in \autoref{sec:  discussion}.
In this paper, we adopt a flat-$\Lambda$CDM cosmology, with parameters $H_0 = 70$ km s$^{-1}$ Mpc$^{-1}$, $\Omega_\Lambda = 0.7$ and $\Omega_m = 0.3$.

\section{The Field of PKS 1610-771} \label{sec: data}
\subsection{ASKAP Detection of 21-cm absorption}
The observations of PKS 1610-771 were part of a FLASH pilot study towards bright ($20$ GHz flux density above $0.5$ Jy) and compact radio sources \citep{Sadler2020} without pre-selection based on criteria such as \ion{Mg}{ii} absorption.
The $21$-cm \ion{H}{i} absorber was first detected with data taken using the six-antenna Boolardy Engineering Test Array \citep[ASKAP-BETA;][]{Hotan2014, McConnell2016} in the period July 2014 to February 2016.
During commissioning time with the ASKAP Early Science array \citep[ASKAP-12;][]{Hotan2021}, the absorber was re-observed with greater sensitivity (rms noise $12.7$ mJy per channel compared to $16.5$ mJy per channel for BETA) in January to February 2017.
In total, PKS 1610-771 was observed for 7 hours using ASKAP-BETA and ASKAP-12. 
For both observations, the frequency resolution was $18.5$ kHz, corresponding to a velocity resolution of $5.7$ km s$^{-1}$ in the absorber rest frame.

%%% Figure 1
\begin{figure}
    \includegraphics[width=\columnwidth]{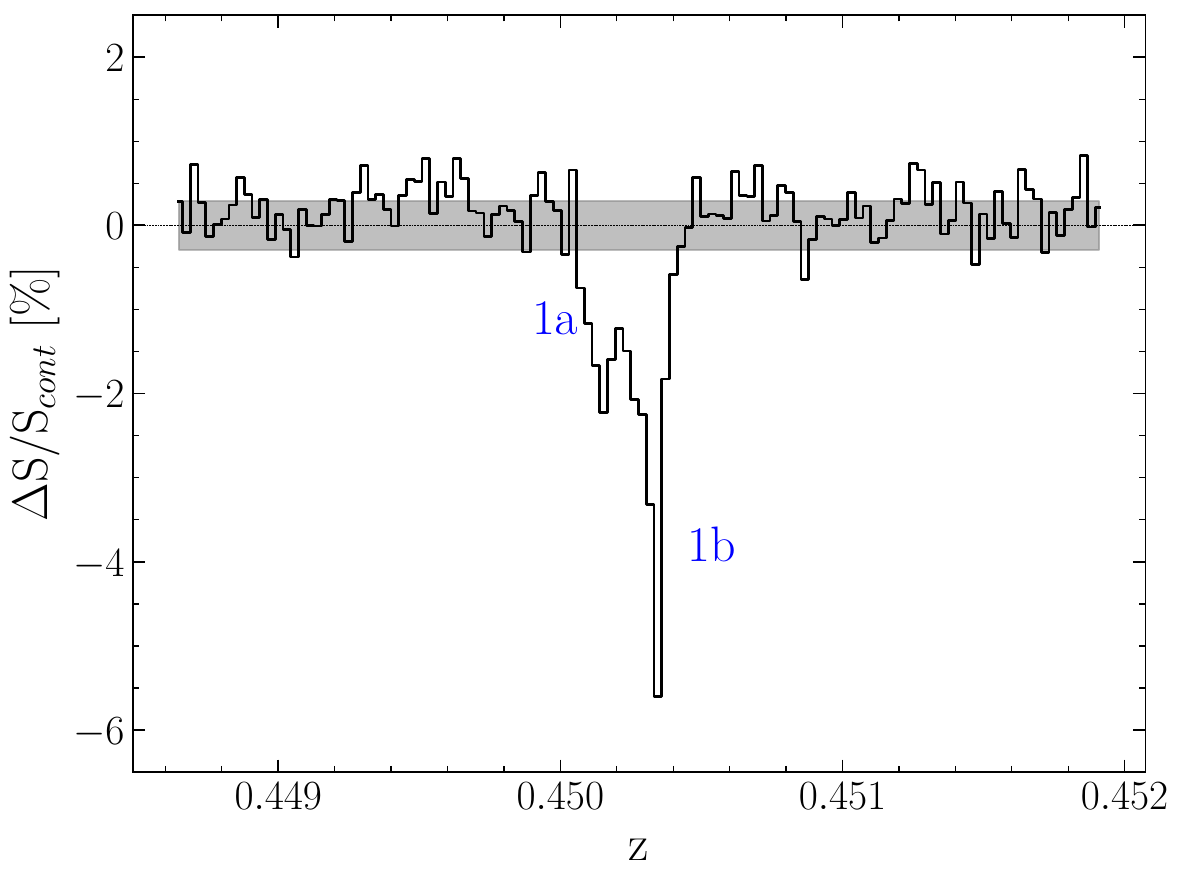}
    \caption{
    21-cm \ion{H}{i} absorption-line profile along the line of sight to the bright radio source PKS\,1610-771. 
    The light grey band indicates the 1$\sigma$ limit in optical depth.
    There are two distinct velocity components (labelled 1a and 1b) that are $30\ \text{km s}^{-1}$ apart. 
    The spectrum is reproduced from \citet{Sadler2020}.}
    \label{fig:HI_1610}
\end{figure}

The $21$-cm absorber detected towards PKS 1610-771 is at redshift $z_{\rm abs} = 0.4503$.
As can be seen from Fig. \ref{fig:HI_1610} (reproduced from \citet{Sadler2020} here for clarity), there are two components $30\ \text{km s}^{-1}$ apart in radial velocity.
Very-long-baseline interferometry (VLBI) imaging of the quasar at $8.4$ GHz \citep{Ojha2010} reveals a source size of around $5$ mas, equivalent to roughly $30$ pc at $z_{\rm abs} = 0.4503$.
In calculating the column density of the neutral hydrogen, we assume a covering factor $f = 1$, because the projected size of the background source at the absorber redshift is smaller than the typical size of \ion{H}{i} clouds \citep[$\gtrsim 100$ pc, ][]{Braun2012}.
The resulting \ion{H}{i} column density, assuming the fiducial spin temperature of $100$ K, is $N_{\ion{H}{i}} = (2.7 \pm 0.1) \times 10^{20} \cdot \rm{[T_{s}/100 K]} \cdot \rm{atoms\ cm}^{-2}$.
As the optical depth of the $21$-cm line is inversely proportional to spin temperature, and the harmonic mean spin temperature of galactic \ion{H}{i} is $\sim 300$ K \citep{Murray2018}, we expect the absorber to be a DLA ($N_{\ion{H}{i}} > 2 \times 10^{20} \ \text{atoms cm}^{-2}$).
We use $\rm{T_{s}} = 100$ K because this traces \ion{H}{i} gas in the cold neutral medium that is likely to collapse into $\rm{H_2}$, and then form stars.
We tabulate the redshifts and column densities of the individual Gaussian components \citep{Sadler2020} required to fit the absorption profile returned from a Bayesian detection method \citep{Allison2012} in \autoref{tab:HI components}.

\begin{table}
	\centering
	\caption{Model parameters derived from fitting \ion{H}{i} absorption towards PKS 1610-771 \citep{Sadler2020}. Column 1 gives the Gaussian component corresponding to Fig. \ref{fig:HI_1610}; column 2 the redshift; column 3 the $\log$ of the \ion{H}{i} column density $(T_{\rm s}=100K, f=1)$; column 4 the velocity FWHM for the optical depth; column 5 the peak component depth normalised by the continuum flux density.}
	\label{tab:HI components}
	\begin{threeparttable}
	\begin{tabular}{ccccc}
		\hline
		ID & $z$ & $\log_{10} N_{\ion{H}{i}}$ & dv & $(\Delta S/S_{\rm cont})_{\rm peak} $ \\
		& & [$\mathrm{atoms\ cm}^{-2}]$ & $[\text{km s}^{-1}]$ & \% \\
		$(1)$ & $(2)$ & $(3)$ & $(4)$ & $(5)$ \\
		\hline
%		\hline
		1a & $0.450184_{-0.000014}^{+0.000014}$ & $20.3_{-0.6}^{+0.6}$ & $47_{-6 }^{+5}$ & $0.0194_{-0.0012}^{+0.0012}$ \\
		& \\
		1b & $0.4503276_{-0.0000015 }^{+0.0000016}$\tnote{}  & $20.0_{-0.5}^{+0.6}$ & $11.2_{-1.1 }^{+1.5}$ & $0.045_{-0.003 }^{+0.003}$ \\
		\hline
	\end{tabular}
	\begin{tablenotes}\footnotesize
		\item[] The $1 \sigma$ errors for each parameter are determined from the marginal posterior distributions calculated using a Monte Carlo Nested Sampling algorithm \citep{Allison2012}, and are particularly small due to the high signal-to-noise ratio (S/N) of these observations.
	\end{tablenotes}
	\end{threeparttable}
\end{table}

\subsection{Ancillary Observations of the PKS\,1610-771 field}
The quasar PKS 1610-771 was first observed by \citet{Courbin1997} with the New Technology Telescope (NTT) in 1995 to confirm the nature of a gravitationally lensed quasar candidate.
Imaging centred on the quasar revealed four galaxy-like objects (referred to as objects A, B, C \& D) within a few arcseconds of the quasar, with object D only visible after a 2D PSF subtraction of the QSO light \citep[see Figure 1 in][]{Courbin1997}.

Early spectroscopic observations of PKS\,1610-771 reveal that this quasar is highly reddened \citep{Hunstead1980, Courbin1997}, implying that there is significant dust absorption in the host galaxy of the quasar or in intervening galaxies along the line of sight. 
Such a reddened QSO may be excluded from optical DLA samples \citep{Krogager2019}, which highlights how $21$-cm \ion{H}{i} surveys without pre-selection can overcome selection biases related to dust. 

More recent spectroscopic observations of objects A (impact parameter 8.8\,kpc) and B (impact parameter 17.8\,kpc) with the $8$-m Gemini-South telescope revealed them to be galaxies whose redshifts fell within $100$ km s$^{-1}$ of the neutral gas detected by ASKAP \citep{Sadler2020}.
While this suggests that the \ion{H}{i} absorption is likely to be associated with one of these galaxies, the exact nature and origin of the \ion{H}{i} gas seen in absorption remained unclear.

\section{MUSE Observations of the PKS 1610-771 field}
\label{sec: MUSE_data}
Of the two new intervening detections from the pilot study, only the absorber towards quasar PKS 1610-771 had candidate host galaxies near redshift $z_{\rm abs} = 0.4503$ \citep{Sadler2020}.
Follow-up MUSE observations were taken centred on PKS 1610-771 for a total of three hours on source.
They were carried out in service mode (programme 0103.A-0656, PI: E. Sadler) in three separate `observing blocks' (OBs) on the nights of 2019 April 12, May 8 and 9.
Each observing block was divided into two sub-exposures (T\textsubscript{exp}=$2 \times 1800$ s) with a $\ang{90}$ rotation and a sub-arcsec dithering pattern was applied between these to minimise artefacts and obtain more uniform noise properties in the dataset.
The field-of-view was $60\ \mathrm{arcsec} \times 60\ \mathrm{arcsec}$ with a $0.2$ arcsec/pixel scale using the Wide Field Mode and the instrument's `nominal mode' used has a spectral coverage spanning $4800-9300$ \AA.
At the redshift of the intervening \ion{H}{i} gas $z_{\rm abs}=0.4503$, this covers strong emission lines from [\ion{O}{ii}] $\lambdaup\lambdaup\, 3727, 3729$ to [\ion{O}{iii}] $\lambdaup\lambdaup\, 4959, 5007$.
The GALACSI Adaptive Optics (AO) system was used to improve the seeing.
This AO system consists of four artificial sodium Laser Guide-Stars to correct for atmospheric turbulence at the cost of blocking $\sim 200$ \AA \ centred around the rest-frame \ion{Na}{i} D line to prevent contamination and saturation of the detector.
Consequently, the H$\delta\, \lambdaup\, 4102$ line at the DLA redshift is not covered, but stronger Balmer lines (H$\beta\, \lambdaup\, 4861$ and H$\gamma\, \lambdaup\, 4340$) are available in the wavelength coverage.

The raw MUSE exposures were reduced using version 2.6.2 of the ESO MUSE pipeline and associated static calibrations \citep{Weilbacher2016}.
Each raw exposure was corrected using master bias, flat-field and arc lamp exposures based on data taken closest in time to the science observations.
The MUSE line-spread functions (LSF) part of the pipeline package were used as the parameters of the LSF are considered stable.
The raw exposures were then processed with the \textsc{scibasic} recipe, using the above calibrations to remove the instrument signature.
During the removal of the sky background with \textsc{scipost}, corrections for the Raman scattered light from the lasers were made and a barycentric reference was adopted to make sure the wavelength calibration was consistent with the ASKAP data.
These individual exposures were then aligned using \textsc{exp\_align} to ensure accurate astrometry and then finally combined using the \textsc{exp\_combine} recipe.

The data reduction pipeline (v 2.6.2) for MUSE has sub-optimal sky subtraction that leaves artifacts in the final product.
It is essential to remove these sky residuals for the detection of emission-line objects without detectable continuum and later, accurate measurement of emission-line fluxes.
The principal component analysis (PCA) method from \citet{Husemann2016} was found to be effective and it significantly improved the final sky subtraction.
PCA components were created for regions of sky selected by the user, and then applied to the data cube. 
These minimalised sky residuals, particularly around the bright night sky emission lines at $5577$ and $6300$ \AA, and the resulting spectra are presented here.
The known wavelengths of the OH night sky emission lines were used to check the wavelength calibration by turning off the sky subtraction for one exposure.
Using this method, the wavelength solution was found to be accurate to $20\ \text{km s}^{-1}$.
The resulting point spread function (PSF) measured from bright sources near the field centre using a Gaussian profile has a FWHM of $0.68$ arcsec at $7000$ \AA.
Measured fluxes are estimated to have an uncertainty of $\pm 30$\% after comparing \textit{R} band magnitudes of PKS1610-771 and nearby stars with values found in the literature.

\section{Analysis and Results}
\label{sec: analysis}

\subsection{Associated Galaxies}

The \textsc{ProFound}\footnote{github.com/asgr/ProFound} algorithm \citep{Robotham2018} is used to identify continuum sources in the field.
There are several misidentifications due to various overlapping sources such as the QSO PKS 1610-771 and galaxy A, which are corrected for manually using \textit{profoundSegimFix} \citep{Bellstedt20, Foster21}.
To search for objects without detectable continuum, the MUSE Line Emission Tracker (\textsc{MUSELET}) module of the \textsc{MPDAF}\footnote{https://mpdaf.readthedocs.io/en/latest/index.html} package \citep{MPDAF} is used to systematically search for emission-line galaxies in the field.
Combined with \textsc{ProFound} and a visual inspection, a complete search for associated galaxies was performed down to the detection limit of $\sim 25.5$ mag (in the $R$ band) and flux limit of $3 \times 10^{-18}\ \mathrm{erg\ s}^{-1}\ \mathrm{cm}^{-2}$ at $7000$ \AA. 
This corresponds to a stellar mass limit of $\log(M_{*}/M_{\odot}) \sim 8$ and dust-uncorrected star formation rate of $0.02 \ \text{M}_{\odot}\ \text{yr}^{-1}$ at the absorber redshift $z_{\rm abs} = 0.4503$.
Galaxies with stellar masses and SFRs below these values will not be detected in the MUSE data, and it is possible that small, passive galaxies at the absorber redshift are missed during source finding.

\begin{figure}
    \includegraphics[width=\columnwidth]{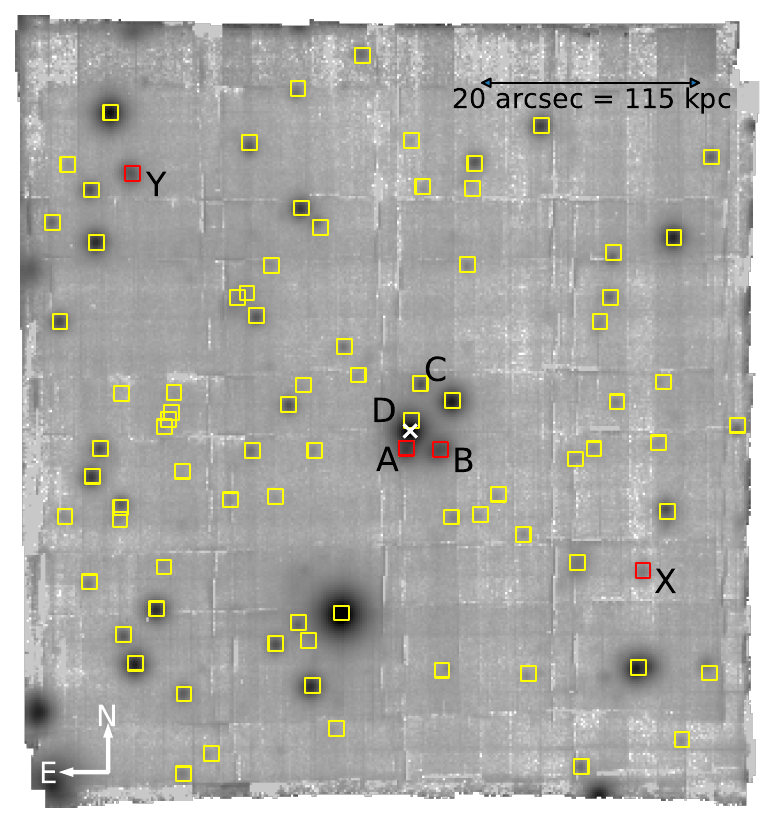}
    \caption{Overlaid on a white-light image of the field are all identified objects, with the quasar PKS 1610-771 at the centre marked by a white cross. 
    Galaxies detected within $1000$ km s$^{-1}$ of the \ion{H}{i} $21$-cm absorber have been marked in red, while other objects in the field are marked in yellow.
  Sources near the edge of the field are not identified. 
  North is up and east is left. }
    \label{fig:Whitelight}
\end{figure}

Spectra for continuum sources are extracted using 1 arcsecond circular apertures centred on the flux-weighted centres determined by \textsc{ProFound} for each object.
Custom apertures for objects near the bright QSO (galaxies A and B) are created to minimalise flux contamination when generating the spectra.
For emission objects detected by MUSELET, a 0.5 arcsecond radius aperture is used to extract spectra.
Redshifts are obtained using the \textsc{4XP}\footnote{https://github.com/lukejdavies/FourXP} (Davies et al. in prep.) package in R.
\textsc{4XP} is in development for the 4-metre Multi-Object Spectroscopic Telescope (4MOST) extragalactic redshifting pipeline and is based on the spectral cross-correlation program \textsc{AutoZ} \citep{Baldry2014}.
The spectra are also visually inspected to search for objects at the absorber redshift.
The resulting white-light image in Fig. \ref{fig:Whitelight} contains $84$ marked sources, of which only the four marked in red have measured redshifts within $\pm 1000$ km s$^{-1}$ of the absorber at $z_{\rm abs} = 0.4503$.
These four galaxies form a possible galaxy group with velocity dispersion $190$ km s$^{-1}$.

Objects C and D \citep{Courbin1997} detected at low angular separation to the quasar are found not to be associated with the absorber.
Object C is identified as a faint M-type star from its prominent TiO bands (see Fig. \ref{fig:galC} for the spectrum).
Object D is obscured by the point spread function of the bright quasar. 
Horizontal and vertical slices across the expected position of galaxy D are initially used to determine candidate emission lines by visually inspecting changes in the spectrum of the quasar.
Then, a spectral PSF subtraction \citep{Hamanowicz2020} is used to reveal these obscured lines.
We note that this method does not allow recovery of the object's continuum.
Potential emission lines are found near $5587$ and $7290$ \AA \ (see Fig. \ref{fig:Gal_D_lines}), corresponding to [\ion{O}{ii}] and H$\beta$ at $z = 0.5001$.
Although the possible [\ion{O}{ii}] line is very near the $5577$ \AA \ sky emission line, \citet{Hunstead1980} observed a similar emission feature near $5590$ \AA \ four decades prior.
The H$\beta$ emission at $7290$ \AA \ is well-supported by the shifting centroid of the line across spaxels, signifying rotation of the ionised gas, and this places galaxy D near $z = 0.5$.
We note that upon inspection of individual exposures, these lines do not consistently appear.
Regardless, there is no evident emission from strong lines such as [\ion{O}{ii}], [\ion{O}{iii}] and H$\beta$ near $z_{\rm abs} = 0.4503$ in front of the quasar, and so object D is likely to be a background object to the $z=0.45$ galaxies. 
\begin{table*}
	\centering
	\caption{Position of galaxies with redshifts within $1000$ $\text{km s}^{-1}$ of the DLA. Column 1 gives the ID marked in the white-light image of Fig. \ref{fig:Whitelight}; column 2 the right ascension; column 3 the declination; column 4 the angular separation in arcsec; column 5 the impact parameter in kpc; column 6 the redshift; column 7 the redshift error; column 8 the velocity with respect to the peak of the \ion{H}{i} absorber; column 9 the absolute $r$-band magnitude.}
	\label{tab:Positions}
	\begin{tabular}{cccccccccc}
		\hline
		Galaxy & RA & DEC & $\delta$ & $b$ & $z$ & $\Delta z$ & $v_{\rm DLA}$ & $M_{r}$  \\
		& \multicolumn{2}{c}{($\mathrm{J}2000$)}  & (arcsec) &  (kpc) & & & $\text{km s}^{-1}$ & mag \\
		$(1)$ & $(2)$ & $(3)$ & $(4)$ & $(5)$ & $(6)$ & $(7)$ & $(8)$ & $(9)$ \\
		\hline
		A & 16:17:48.29 & -77:17:24.72 & 1.53 & 8.82 & 0.45061 & 0.00008 & $60 \pm 20$ & -21.2\\
%		\hline
		B & 16:17:47.40 & -77:17:24.81  & 3.08 & 17.8 &  0.45038 & 0.00007 &  $20 \pm 20$ & -21.1 \\
%		\hline
		X & 16:17:42.06 & -77:17:35.34  & 23.6 & 136 & 0.45129 & 0.00008 & $200 \pm 20$ & -17.9\\
%		\hline
		Y & 16:17:55.52 & -77:17:00.83 & 33.0 & 190 & 0.45279 & 0.00009 & $520 \pm 30$ & -20.5\\
		\hline
	\end{tabular}
\end{table*}

\begin{figure*}
    \centering
    \includegraphics[width=.31\linewidth]{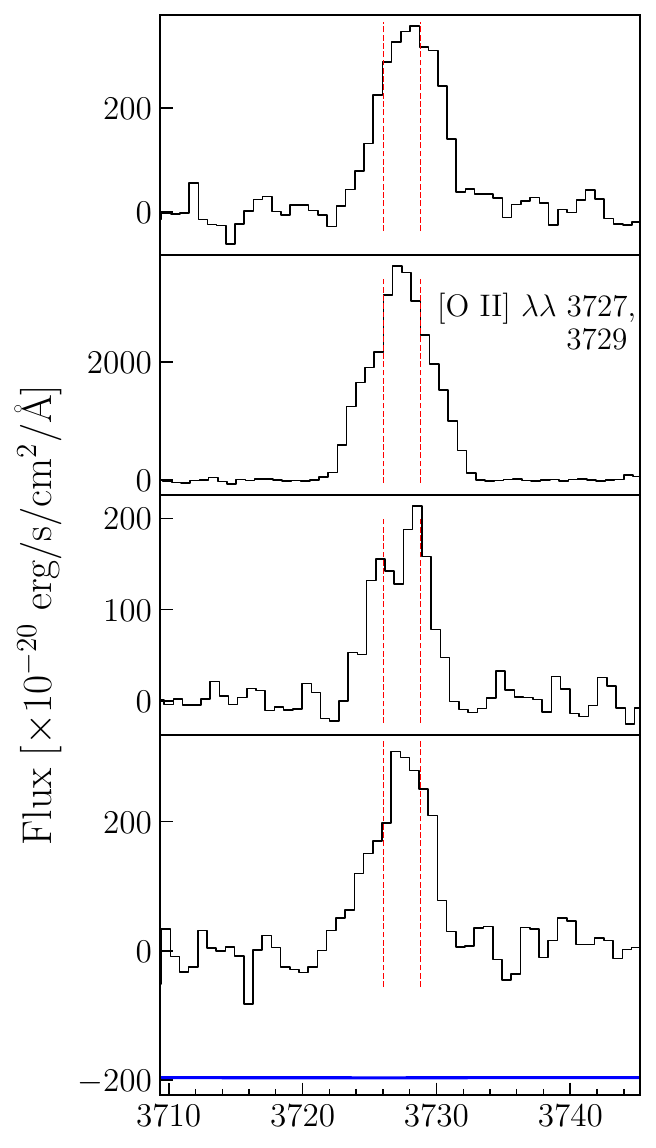}\hfill
     \raisebox{-0.43cm}{\includegraphics[width=.28\linewidth]{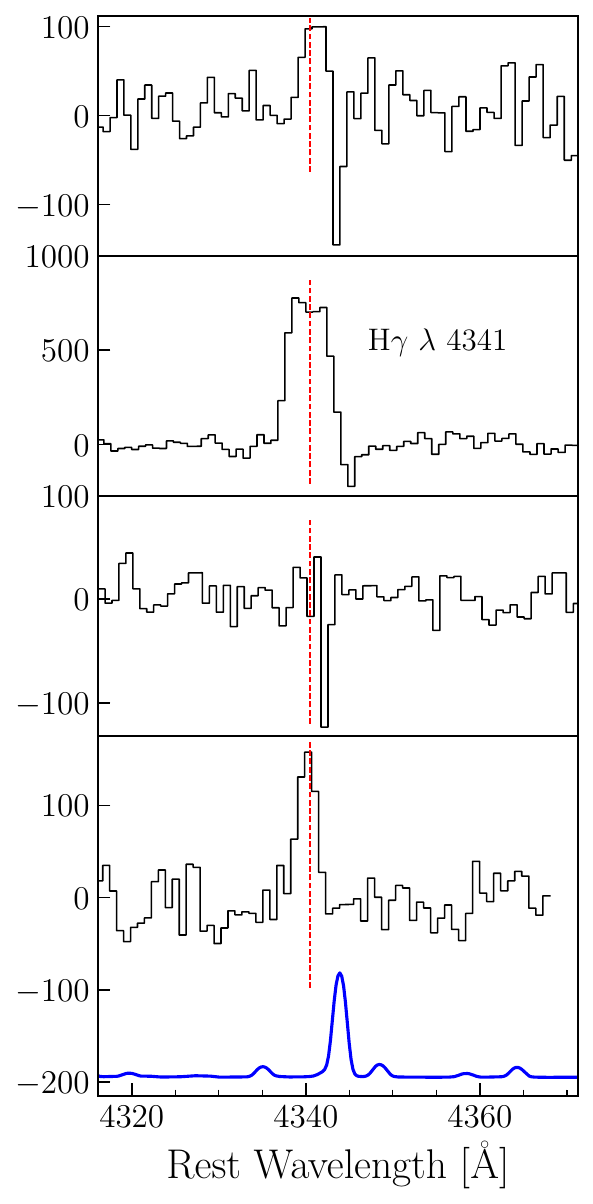}}\hfill
    \includegraphics[width=.393\linewidth]{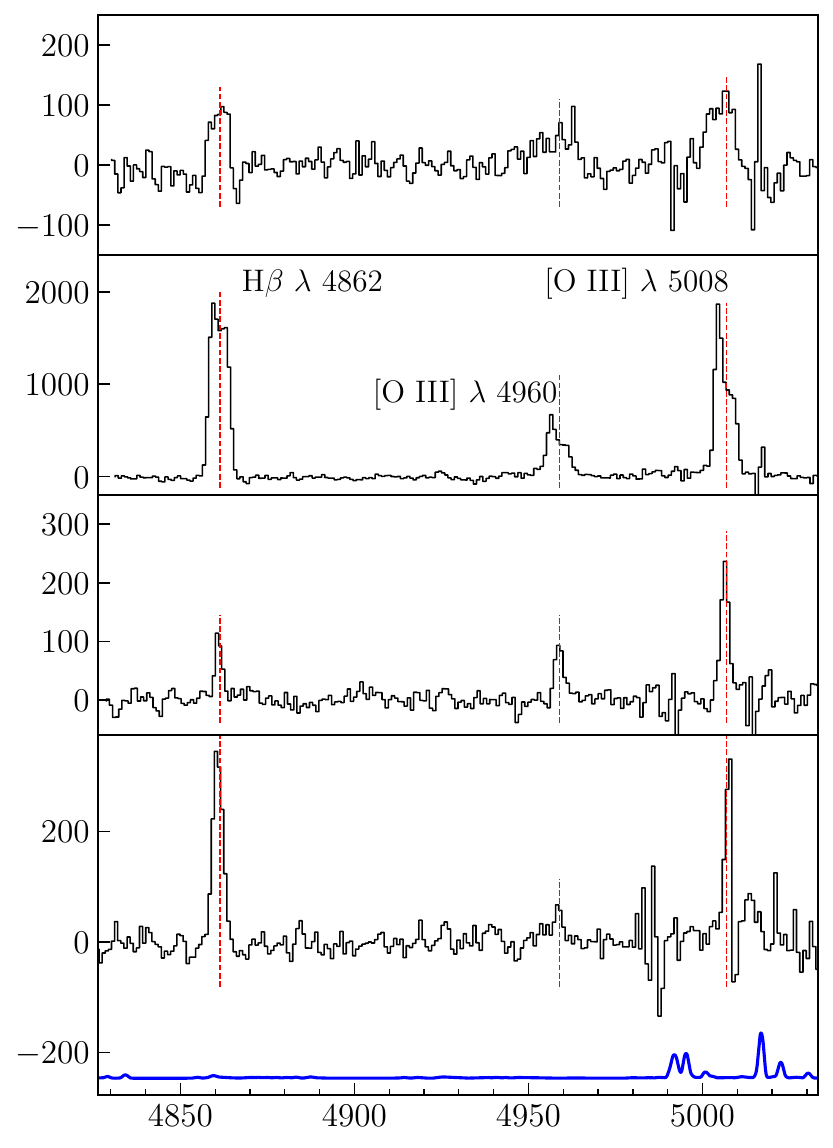}
    \caption{Emission lines for galaxies A, B, X and Y from top to bottom plotted in the rest frame. 
     		Included are [\ion{O}{ii}], H$\gamma$, H$\beta$ and [\ion{O}{iii}], with expected positions marked by red vertical lines}.
     		The continuum-subtracted spectrum is in black, with the vertically offset and arbitrarily scaled sky spectrum in blue (bottom panel).
     		Residuals from the subtraction of night sky lines near $6300$ \AA \ and $7200-7300$ \AA \ affect integrated flux measurements for the H$\gamma$ and [\ion{O}{iii}] doublet emission lines.
     \label{fig:emission_fluxes}
    \label{fig:my_label}
\end{figure*}

\begin{figure*}
	\includegraphics[width=\linewidth]{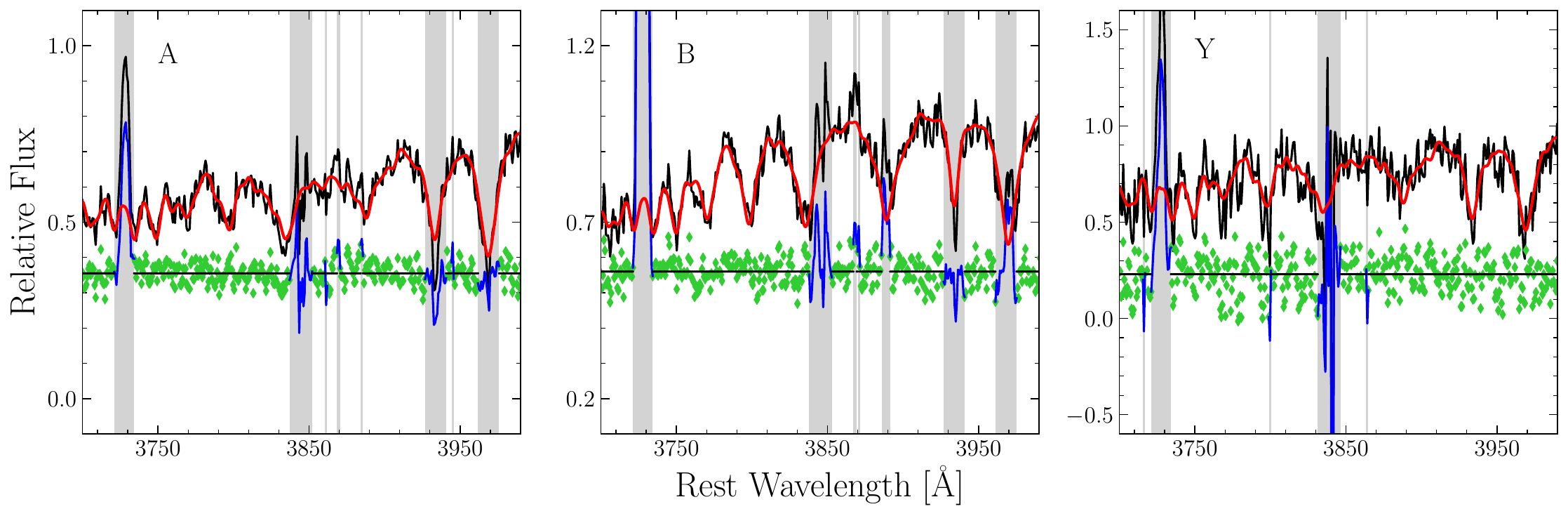}
     \caption{Spectra of higher-order Balmer lines for galaxies A (left), B (centre) and Y (right) in the rest frame. The original spectrum is in black with template fits from \textsc{ppxf} overlaid in red. The grey shaded areas indicate regions that have been masked during fitting -- these correspond to emission lines (e.g. [\ion{O}{ii}]), the $5577$ \AA \ sky line and other regions with strong sky residuals. In green are the residuals that are vertically offset from the black horizontal line and the masked residuals are in blue.}
	\label{fig:balmer_lines}
 \end{figure*}

The positions and redshifts of the galaxies associated with the DLA at $z_{\rm abs} = 0.4503$ are tabulated in \autoref{tab:Positions}.
For galaxies A and B, the continuum emission is clear and there is evident \ion{Ca}{ii} H\&K and \ion{Na}{i} D absorption.
The remaining galaxies (X and Y) have faint continuum emission, but \ion{Ca}{ii} absorption is still present.
Emission lines [\ion{O}{ii}] and H$\beta$ can be seen in all four galaxies (Fig. \ref{fig:emission_fluxes}), with H$\gamma$ also prominent in galaxies B and Y (see \ref{fig:OII_NB} for a synthetic continuum-subtracted narrowband image centred on [\ion{O}{ii}]).
Fig. \ref{fig:balmer_lines} shows that the spectra of galaxies A, B and Y also feature higher-order Balmer lines ($\mathrm{H}\epsilon$, $\mathrm{H}\zeta$, $\mathrm{H}9$, $\mathrm{H}10$, $\mathrm{H}11$ \text{and} $\mathrm{H}12$).
While analysis of the underlying stellar population for these galaxies is difficult due to the low signal-to-noise ratio of the continuum, the lines reveal the presence of a significant population of A- or F-type stars.
Indeed, this is supported by fits using the Penalized Pixel-Fitting software \textsc{PPXF}\footnote{https://pypi.org/project/ppxf/} v7.0.0 \citep{Cappellari2004, Cappellari2017} and MILES stellar library \citep{Sanchez-Blazquez2006, Falcon-Barroso2011}, where A-type stars are among the highest-weighted stellar templates for the three galaxies.

\begin{table*}
	\centering
	\caption{Emission line fluxes and star formation rates for galaxies A, B, X and Y. Column 1 gives the galaxy ID; column 2 the [\ion{O}{ii}] flux; column 3 the $\mathrm{H}\gamma$ flux; column 4 the $\mathrm{H}\beta$ flux; columns 5 and 6 the [\ion{O}{iii}] $4960$ \AA \ and $5008$ \AA \ fluxes respectively; column 7 the dust-uncorrected star formation rate.}
	\label{tab:Emission Fluxes}
	\small
	\begin{threeparttable}
	\begin{tabular}{ccccccc}
		\hline
		Galaxy & $F(\text{[\ion{O}{ii}]})$ & $F(\mathrm{H}\gamma)$ & $F(\mathrm{H}\beta)$  & $F(\text{[\ion{O}{iii}]})$  & $F(\text{[\ion{O}{iii}]})$ & $\mathrm{SFR}$ \\
		& $(\mathrm{erg\ s}^{-1}\ \mathrm{cm}^{-2})$ & $(\mathrm{erg\ s}^{-1}\ \mathrm{cm}^{-2})$ & $(\mathrm{erg\ s}^{-1}\ \mathrm{cm}^{-2})$ &  $(\mathrm{erg\ s}^{-1}\ \mathrm{cm}^{-2})$ & $(\mathrm{erg\ s}^{-1}\ \mathrm{cm}^{-2})$ & $\text{M}_{\odot}\ \text{yr}^{-1}$ \\
		$(1)$ & $(2)$ & $(3)$ & $(4)$ & $(5)$ & $(6)$ & (7) \\
		\hline
		A & $(5.1 \pm 1.7) \times 10^{-17} $ &  $< 0.2 \times 10^{-17}$ &  $(1.1 \pm 0.2) \times 10^{-17} $& $< 0.3 \times 10^{-17}$ & $(1.8 \pm 0.4) \times 10^{-17} $ & $0.56 \pm 0.18$  \\
%		\hline
		B & $(52 \pm 8) \times 10^{-17} $ & $(9.3 \pm 2.0) \times 10^{-17} $ & $(23 \pm 4) \times 10^{-17}$ &  $(7.2 \pm 2.0) \times 10^{-17} $ & $(17 \pm 4) \times 10^{-17} $ & $3.4 \pm 1.1$ \\
%		\hline
		X & $(2.4 \pm 1.0) \times 10^{-17} $ &  $< 0.2 \times 10^{-17}$ & $(5.8 \pm 3.0) \times 10^{-18} $ & $(6.4 \pm 3.0) \times 10^{-18} $ & $(1.5 \pm 0.6) \times 10^{-17} $ & $0.25 \pm 0.11$ \\
%		\hline
		Y & $(4.1 \pm 1.7) \times 10^{-17} $ & $(9.1 \pm 5.0) \times 10^{-18} $ & $(2.5 \pm 1.1) \times 10^{-17} $ & $< 0.3 \times 10^{-17}$ & $(1.6 \pm 1.2) \times 10^{-17} $ &  $0.28 \pm 0.18$\\
		\hline
	\end{tabular}
	\end{threeparttable}
\end{table*}

\subsection{Star Formation Rates}
Fluxes for the dominant emission lines [\ion{O}{ii}], H$\beta$, H$\gamma$ and [\ion{O}{iii}] are determined using the 1D Gaussian fitting tool in the \textsc{MPDAF} module.
The fits were performed on continuum-subtracted spectra made using \textsc{PPXF}.
This helped obtain more accurate flux values by accounting for the stellar absorption near H$\beta$ and higher-order Balmer lines near [\ion{O}{ii}]. 
The flux values are tabulated in \autoref{tab:Emission Fluxes}.
We estimate the star formation rate for galaxies A, B, X and Y using the empirical relation from \citet{Kennicutt1998}:
\begin{equation}
    \text{SFR}\ = (1.4 \pm 0.4) \times 10^{-41}\  L([\text{{\ion{O}{ii}}}])\ \text{M}_{\odot}\ \text{yr}^{-1}.
\end{equation}
This calibration assumes a Salpeter IMF, which we adopt for consistent comparisons with the literature.
These SFR values have not been corrected for dust extinction as H$\alpha$ lies outside the MUSE wavelength range at $z_{\rm abs} = 0.4503$. 
Hence, the H$\alpha /$H$\beta$ Balmer decrement cannot be calculated.
We find that galaxy B is a star-forming galaxy at $z = 0.45$ with $\text{SFR} = 3.4\ \rm{M_{\odot}\ yr^{-1}}$, and its position within the star-forming region of the blue-BPT diagram \citep{Lamareille2010} suggests flux contamination from an active galactic nucleus (AGN) is unlikely.
Galaxies A, X and Y are less active but without dust corrections, it is difficult to decisively label these three galaxies as passive, because the dust-corrected SFR may be larger.
In addition, we cannot determine whether galaxies A, B, X and Y are quiescent or main-sequence galaxies using the $\rm{SFR}-\rm{M}_{*}$ main sequence \citep{Schreiber2015}, because stellar masses cannot be accurately estimated from the limited photometry information available. 

\subsection{Gas Kinematics}
\subsubsection{Neutral Gas}
\begin{figure}
  {\includegraphics[width=\columnwidth]
  {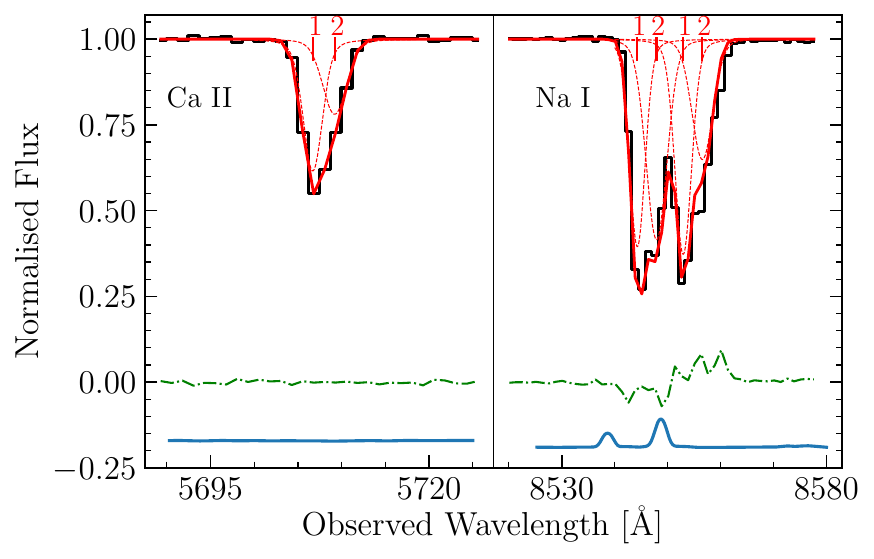}}
  \caption{Normalised MUSE spectrum of \ion{Ca}{ii} K (left) and \ion{Na}{i} doublet (right) in front of QSO PKS 1610-771. The spectrum is plotted in black with Voigt fits in red (thin, dashed lines represent individual components) and residuals in green. There are two components for both ions, most clear in the \ion{Na}{i} fit. The first component coincides with the redshift from ASKAP near $z_{\rm abs} = 0.4503$, while the weaker component is at a redder wavelength. The scaled and offset sky spectrum is in blue.}
  \label{fig:Ca_Na_profiles}
\end{figure}

In the MUSE spectrum of QSO PKS 1610-771 at $z = 1.71$, there are various metal absorption lines.
A closer inspection reveals that these lines belong to two separate systems at redshifts $z \sim 0.45$ and $z \sim 1.61$ (see Fig. \ref{fig:QSO_spec}).
We focus on the $z \sim 0.45$ system consisting of low ionisation potential metal lines \ion{Na}{i} D$_1$ \& D$_2$ and \ion{Ca}{ii} H \& K as the DLA is found at $z_{\rm abs} = 0.4503$.
\textsc{VPFIT} v10.4 \citep{VPFit} is used to fit the doublets after converting the MUSE spectrum to vacuum wavelengths, and two components are found to be necessary for the line profile.
This is in agreement with a visual inspection of the line profiles, where there is an extended wing in the \ion{Ca}{ii} K absorption feature and the \ion{Na}{i} D doublet appears double-peaked.
The separate components and final fits are depicted in Fig. \ref{fig:Ca_Na_profiles}.

In the intergalactic environment of the Milky Way, these ions generally trace extraplanar high-velocity clouds with \ion{H}{i} column densities ranging from $10^{17}$ to  $10^{20}$ atoms cm$^{-2}$ \citep{Richter2005, Bekhti2008}.
Similarly, we find from Voigt profile fits of these ions that the redshift of the higher column density component aligns with the velocity of the neutral gas detected by ASKAP.
Due to the poorer velocity resolution of MUSE, we cannot link the \ion{Na}{i} and \ion{Ca}{ii} detection to a single component found in the ASKAP data (1a or 1b).
There is only a 30 km s$^{-1}$ separation between the Gaussian centroids of components 1a and 1b, which cannot be resolved by the MUSE instrument that has a velocity resolution of $110$ km s$^{-1}$ at $7000$ \AA.
In addition, we see in Fig. \ref{fig:Ca_Na_profiles} that there is a second fitted component, redshifted $110$ km s$^{-1}$ from the former.
An examination of the $21$-cm absorption line in Fig. \ref{fig:HI_1610} reveals there is no corresponding feature near $z = 0.4509$, suggesting the gas probed is of lower \ion{H}{i} column density or higher spin temperature.
Regardless, the MUSE data have uncovered another component to the absorber.

\subsubsection{Ionised Gas}
Absorption-line studies probe pencil-beam sightlines towards the background source and allow statistical studies of the neutral gas amount and distribution at intermediate redshifts \citep[e.g.][]{Sadler2020}.
However, the interpretation of the \ion{H}{i} kinematics requires further optical spectroscopy and imaging of associated galaxies.
To unravel the relationship between the neutral gas and the galaxy overdensity at $z = 0.45$, it becomes essential to analyse the absorber velocity relative to the stellar or ionised gas kinematics of galaxies A, B, X and Y.
The S/N of the stellar continuum is too low ($< 3$) in individual spaxels to measure the line-of-sight velocity distribution (LOSVD) for stars reliably.
Instead, we use \textsc{PPXF} to derive the LOSVD for the ionised gas and generate velocity maps.
The prominent emission lines in each spaxel of the galaxy are constrained to have the same velocity $V$ and dispersion $\sigma$, and are fitted using emission line templates.
Only the emission lines of galaxy B meet the signal-to-noise ratio cutoff $S/N > 3$ for a significant number of individual spaxels, and the remaining galaxies A, X and Y cannot have their ionised gas mapped.

\begin{figure*}
  \centering
      \includegraphics[width=0.9\textwidth]{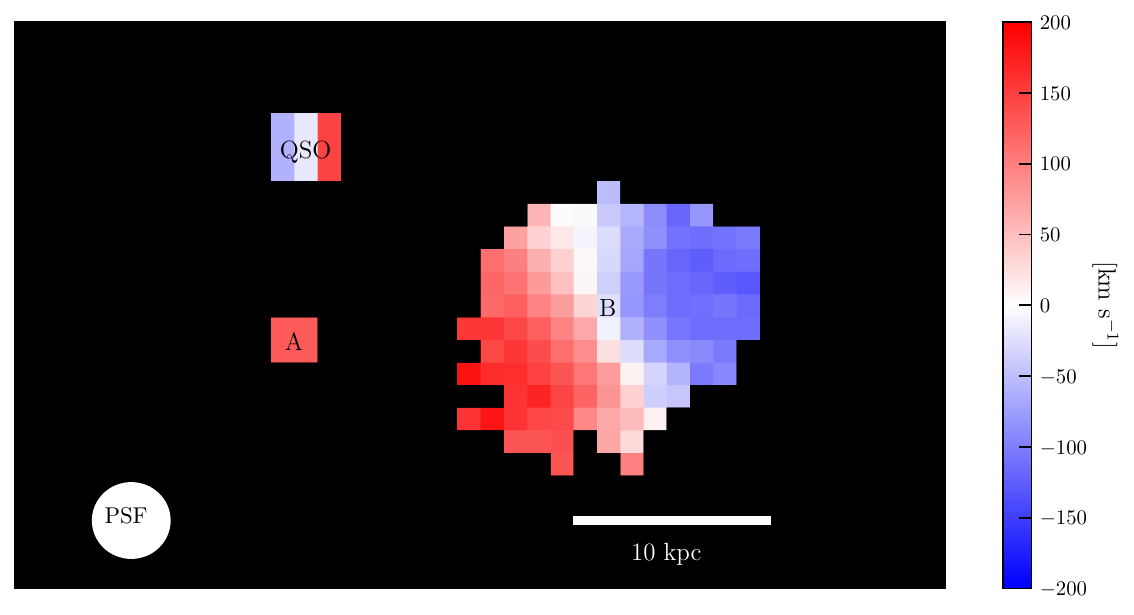}
  \caption{Observed kinematic map of the ionised gas in galaxy B, with zero velocity at its systemic redshift. 
  At the position of the QSO PKS 1610-771, the three vertical stripes correspond to the neutral gas velocities along the quasar sightline. 
  The blue and white stripes reflect the velocities of the two \ion{H}{i} components measured with ASKAP, while the red stripe is from Voigt profile fitting of the \ion{Ca}{ii} and \ion{Na}{i} doublets.
  We note here that these velocities are not spatially resolved and occur from coincident sightlines through the gas.
  The position of galaxy A is marked with a single velocity corresponding to its systemic redshift.
  At the bottom left is a circle with diameter equivalent to the FWHM of the seeing at $7000$ \AA.
  The spatial sampling is 0.2 arcsec/pixel, corresponding to $1.15$ kpc/pixel at $z_{\rm abs} = 0.4503$.}
  \label{fig:Key_map}
\end{figure*}

[\ion{O}{ii}] and H$\beta$ are the emission lines used to create the observed rotational velocity map in Fig. \ref{fig:Key_map} as they are the strongest lines in the spectrum, allowing more accurate measurements in the outskirts of the disc.
All velocities are represented relative to galaxy B's systemic redshift, which is calculated by applying a further correction to the original \textsc{4XP} fit using \textsc{PAFit} \citep{Krajnovic2006}.
This correction to the \textsc{4XP} redshift is useful because we want to compare the kinematics of the neutral gas in front of the quasar with galaxy B's ionised gas. 
As \textsc{4XP} is a spectral cross-correlation software, there are contributions from other components of the spectrum in the final redshift such as the stellar absorption lines.
For comparisons between gas kinematics, the \textsc{PAFit} correction enables us to more accurately measure the zero-point of galaxy B's ionised gas rotation map.
Additionally, \textsc{PAFit} also returns a kinematic position angle (PA) of $114 \pm 5$\degree \ from the fitting of [\ion{O}{ii}] and H$\beta$ rotation curves.
Galaxy A is included in the diagram to illustrate its projected position and relative velocity.
Velocities of the neutral gas are also depicted near the expected position of the absorber for comparison to galaxy B's rotating disc.
These velocities are unresolved as they represent coincident sightlines through the gas.
The PSF FWHM of $0.68$\,arcsec at $7000$ \AA \ is represented in the bottom left of Fig. \ref{fig:Key_map} by the diameter of the circle.

\section{Relationship between the Neutral Gas and Associated Galaxies}
\label{sec: discussion}
In total, we have three kinematic measurements of the neutral gas at $z \sim 0.45$.
Two velocities are derived directly from the components of the initial ASKAP detection (Fig. \ref{fig:HI_1610}), and a third by proxy using the low-ionisation metal absorption lines \ion{Na}{i} and \ion{Ca}{ii} (Fig. \ref{fig:Ca_Na_profiles}).
Due to the overlap in redshift between the ASKAP \ion{H}{i} detection and component `1' from the MUSE spectra, they are likely tracing the same gas.
We henceforth only consider the $21$-cm components because ASKAP has a significantly better velocity resolution of $5.7$ km s$^{-1}$ at $z_{\rm abs} = 0.4503$ compared to $\sim 110$ km s$^{-1}$ at $7000$ \AA \ for MUSE.
If the metal lines are indeed tracing the neutral gas, the distinct components seen in the ASKAP data are possibly not resolved in the MUSE data and hence, appear as a single component.
The three components will be henceforth referred to as 1a, 1b and 2 in order of increasing redshift. 
Their properties are listed in \autoref{tab:Components Summary}.

\begin{table}
	\centering
	\caption{Properties of the three gas components considered. Components 1a and 1b are taken from the ASKAP detection, while component 2 at higher redshift is obtained from fitting low-ionisation ions in the MUSE spectrum of the quasar. Column 1 is the component ID; column 2 the redshift of the component; columns 3 and 4 the velocities of the component with respect to the systemic redshift of galaxies A and B respectively; column 5 the absorption lines associated with each component.}
	\label{tab:Components Summary}
	\begin{threeparttable}
	\begin{tabular}{ccccc}
		\hline
		Component & $z$ & $v_{\rm A}$ & $v_{\rm B}$ & Lines \\
		& & [$\text{km s}^{-1}$] & [$\text{km s}^{-1}$] &  \\
		$(1)$ & $(2)$ & $(3)$ & $(4)$ & $(5)$  \\
		\hline
		 1a  & $0.45018$ & $-90$ & $-40$ & \ion{H}{i}, \ion{Ca}{ii}, \ion{Na}{i}\tnote{a} \\
%		\hline
		 1b & $0.45033$ & $-60$ & $-10$ & \ion{H}{i}, \ion{Ca}{ii}, \ion{Na}{i}  \\
%		\hline
		2 & $0.4509$ & $+60$ & $+110$ &  \ion{Ca}{ii}, \ion{Na}{i}\\
		\hline
	\end{tabular}
	\begin{tablenotes}\footnotesize
		\item[a] \ion{Ca}{ii} and \ion{Na}{i} component near $z=0.4503$ in MUSE spectrum could trace both ASKAP \ion{H}{i} components due to insufficient velocity resolution.
	\end{tablenotes}
	\end{threeparttable}
\end{table}

\begin{figure}
\includegraphics[width=0.9\columnwidth]{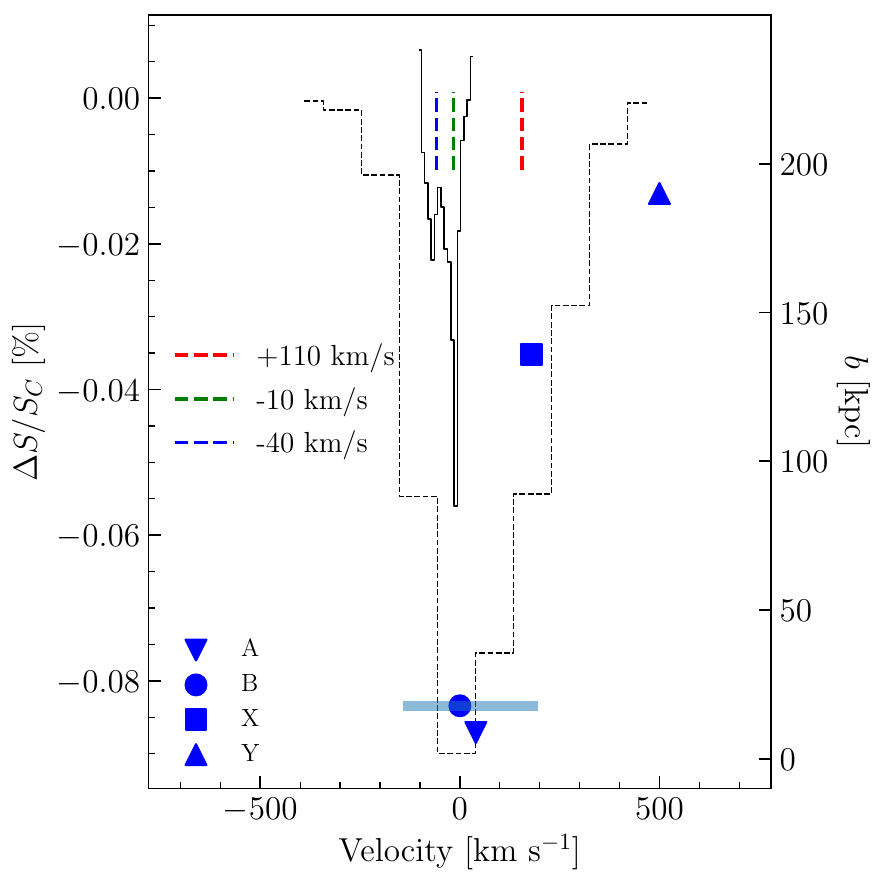}
\caption{The ASKAP \ion{H}{i} absorption line profile in velocity space is plotted in black with the velocities of its two components labelled by the dashed blue (1a) and green (1b) lines.
This spectrum is to scale and plotted against the fraction of absorbed background continuum $\Delta S / S_C$.
  The dashed spectrum is \ion{Ca}{ii} K absorption from MUSE and has its second component marked in red (2).
  Note that the first component of the \ion{Ca}{ii} K metal line aligns with the ASKAP components and is not plotted. 
  Each of the galaxies are represented by blue markers using their respective redshifts and impact parameters to the gas ($b$, right axis).
  At the position of galaxy B, a horizontal line marks the extent of its rotation.
  All velocities are with respect to the systemic redshift of galaxy B ($z = 0.45038$) to be consistent with Fig. \ref{fig:Key_map}.}\label{fig:1D_keymap}
\end{figure}

With the relevant neutral gas velocities identified, and associated galaxies analysed, we can now try to understand the relationship between the gas and surrounding galaxies.
In Fig. \ref{fig:1D_keymap}, the impact parameters and redshifts of galaxies A, B, X and Y are compared to the velocities of the gas seen in absorption by both ASKAP and MUSE.
Galaxies A and B have the smallest projected separations to the absorber and are closest to it in velocity space.
Since the $21$-cm absorber detected with ASKAP has DLA-equivalent column density, there are restrictions on which galaxies are directly related to the neutral gas in front of the quasar.
The characteristic radius of $21$-cm DLA sources is estimated to be less than $20$ kpc from studies at lower redshift \citep{Borthakur2016, Reeves2016, Curran2016, Dutta2017}, constraining the source of the \ion{H}{i} to be at a similar impact parameter.
We hence exclude galaxies X and Y from having a direct relationship with the DLA. 
They have large projected separations ($136$ and $189$ kpc respectively) compared to galaxies A and B, which have impact parameters $< 20$ kpc.
% Component 2, which does not have a corresponding feature in the ASKAP spectrum, may be tracing gas part of the DLA that is of lower column density, higher temperature or perhaps partially ionised.
Further discussion revolves around this pair of galaxies nearest the quasar sightline.

\subsection{Co-rotation with Galaxy B's Disc}
The size of the \ion{H}{i} disc in isolated galaxies is roughly twice the optical diameter \citep[e.g.][]{RaoBriggs1993, Boomsma2008}.
Thus, the absorber is possibly probing gas co-rotating with a galaxy, and this can be tested by extrapolating from the kinematics of the ionised disc determined using \textsc{PPXF}.
Galaxy A's rotation map is difficult to measure because the [\ion{O}{ii}] emission spans $0.8 \times 0.8$ arcseconds, which is marginally larger than the PSF FWHM.
Therefore, our discussion focuses on galaxy B, which has resolved rotation.

From Fig. \ref{fig:Key_map}, we see that the redshifted side of galaxy B is orientated towards the QSO sightline.
Thus, components 1a ($-40\ \text{km s}^{-1}$) and 1b ($-10\ \text{km s}^{-1}$) corresponding to the velocities of the \ion{H}{i} detection from ASKAP are not consistent with galaxy B's rotating disc, even after accounting for uncertainties in redshift measurements.
Component 2 from the \ion{Na}{i} and \ion{Ca}{ii} aligns in velocity with this phenomenon as it is also redshifted with respect to the systemic velocity of galaxy B.
Given that there is no corresponding \ion{H}{i} absorption for component 2, the neutral gas is likely warmer, and may trace gas in the outer disc of galaxies heated by the radiation field \citep{Maloney1993}.
However, other processes are required to account for the kinematics of the ASKAP \ion{H}{i} detection.

\subsection{Outflows}
In current models, outflows are expected to be preferentially aligned with a galaxy's minor axis as enhanced resistance along the galactic plane effectively collimates outflows into a biconical shape \citep{Nelson2019, Peroux2020}.
The azimuthal angle $\alpha$, defined as the angle between the galaxy major axis and projected position of the quasar, is typically used to distinguish between accretion and winds \citep[e.g.][]{Zabl2019, Schroetter2019}.
For galaxy B, it is found that $\alpha = 57$\degree ($i = 47$\degree, PA $= 114$\degree) which indicates that the absorber is preferentially aligned with the minor axis.
Stellar winds are found in emission-line galaxies at intermediate redshifts, \citep{Zhu2015} and the alignment with the minor axis suggests the neutral gas detected is wind material.
These winds are likely not driven by an AGN as the galaxy is classified as star-forming using the blue-BPT diagram \citep{Lamareille2010}.
Galaxy A has too low S/N emission lines in individual spaxels to be modelled using \textsc{PPXF}, and its proximity to the quasar prevents accurate measurements of its photometric position angle.
Galaxies X and Y are unlikely candidates due to their $> 100$ kpc impact parameter to the DLA and large velocity difference of $+ 200$ and $+ 520 \ \text{km s}^{-1}$, respectively, to the absorber redshift.
If outflowing gas is responsible for the absorption features, it is far more likely that the winds originate from galaxy B due to its proximity and higher SFR.
However, we cannot rule out outflowing gas from galaxy A being responsible for the neutral gas detected in absorption due to the lack of kinematic modelling.

The high \ion{H}{i} column densities of components 1a and 1b suggest that cold dense neutral gas is being traced.
While galactic winds have been ubiquitously observed at all redshifts up to $z \sim 3$ \citep{Rupke2018}, an important consideration is whether the outflowing cold gas can survive out to $\sim 20$ kpc from galaxy A or B without breaking down or being heated and ionised.
Simulations predict starburst-driven outflows to be multi-phase, consisting of a hot and fast ionised component entraining colder gas from the interstellar disc \citep{Veilleux2005}.
More recent hydrodynamical simulations suggest mechanisms for the cold gas to accelerate to wind speeds and grow in mass \citep{GronkeOh2018, GronkeOh2020}.
Outflows of low-ionised gas traced using \ion{Mg}{ii} absorption have been seen to extend beyond $100$ kpc \citep[e.g.][]{Schroetter2019}, but this traces gas at temperatures of $\sim 1000$ K.
While the spin temperature of neutral hydrogen is always less than or equal to the kinetic temperature of the gas \citep[e.g.][]{Purcell1954}, a mean harmonic spin temperature of $1000$ K for our absorber requires an order of magnitude increase in the \ion{H}{i} column density ($N_{\ion{H}{i}} = 2.7 \pm 0.1 \times 10^{21}$ atoms cm$^{-2}$) to reproduce the observed absorption line.
These higher spin temperatures and super-DLA column densities are not unusual \citep{Kanekar2014, Curran2019, Allison2021b}, and the possibility that the ASKAP neutral gas detection traces outflowing material remains.

Outflows of cool gas have been observed at higher redshift in other galaxies using \ion{Na}{i} D absorption against background stellar continuum \citep[so-called `down-the-barrel' spectroscopy, e.g.][]{Heckman2000, Cazzoli2016}.
While the detected outflow regions in these studies typically extend $< 10$ kpc from the source, this is more likely caused by the faintness in the stellar continuum at larger radii rather than an intrinsic property of the outflow.
The absorption in this system is not down-the-barrel, but the quasar sightline is only $18$ ($9$) kpc from galaxy B (A).
Hence, component 2, which is traced by \ion{Na}{i} and \ion{Ca}{ii} without corresponding $21$-cm absorption in the ASKAP data, is possibly wind material.

\subsection{Inflows}
Galaxies require replenishment of their gas reservoirs to sustain their star formation rates.
Contrary to outflows, cold gas accreting from dark matter filaments are expected to align preferentially with the galaxy major axis in the form of an extended cold gas disc \citep{Ho2019}.
For galaxy B, its inclination $i = 47$\degree\ and azimuthal angle of $57$\degree \ indicate that the orientation is not favourable for probing an extended gaseous disc.
Further, authors in studies of gas accretion select a single `primary' galaxy most likely associated with the absorber within a search radius \citep{Zabl2019}.
This ensures that inflows are responsible for the absorber.
In our case, both galaxies A and B are located within $20$ kpc and $100 \ \text{km s}^{-1}$ of the absorber, and it becomes impossible to guarantee the absorber is uniquely tracing inflowing material.

\subsection{Extragalactic Gas Clouds}
\label{sec: discuss_interactions}
\ion{H}{i} emission maps of local interacting galaxy pairs and groups reveal extragalactic clouds of gas \citep[e.g.][]{Verdes-Montenegro2001, Lee-Waddell2019}.
This phenomena has also been possibly detected at higher redshift in a radio galaxy interacting with a satellite \citep{Allison2019} using \ion{H}{i} absorption and ALMA observations.
These clouds are typically high column density ($\sim 10^{20}$ atoms cm$^{-2}$) and of similar velocity to nearby galaxies.
Galaxies A and B are separated by $17$ kpc, with only a $60 \ \text{km s}^{-1}$ velocity difference, and Fig. \ref{fig:Key_map} reveals they share the same velocity plane, with the redshifted side of galaxy B extending towards galaxy A.
Additionally, the significant contribution of A-type stellar templates during \textsc{PPXF} fitting is indicative of a significant young stellar population.
This suggests a period of enhanced star formation several $100$ Myrs in the past, most likely induced by interactions between galaxies A and B.
\citet{Klitsch2019} similarly find that absorption-line studies may preferentially select interacting galaxies with a wider distribution of gas, and that the galaxies associated with the absorber have more excited ISMs. 
It becomes possible then that components 1a and 1b seen in absorption are probing extragalactic clouds formed from interactions.

\subsection{Nature of the Gas}
The wealth of information obtained from the ASKAP and MUSE instruments allows us to gain a greater understanding of the gas seen in absorption.
Components 1a and 1b from the initial radio detection cannot be explained by the rotating disc of galaxy B, and the emission lines of nearest group member, galaxy A, are too low S/N to discern ionised gas velocities for comparison with the neutral gas kinematics.
Instead, a more likely scenario is that the $21$-cm detection probes extragalactic gas clouds formed by interactions between galaxies A and B.
While outflows from galaxy B cannot be excluded, it is questionable whether such dense amounts of cold neutral gas will still be found in ejected material roughly $20$ kpc from the source. 
Additional constraints on this scenario can be applied if the DLA metallicity is measured and then compared with the galaxy metallicity.
% The absence of metallicity information also means there is not enough information to conclude that component 2 derived from \ion{Na}{i} and \ion{Ca}{ii} absorption is a result of inflows. 
Finally, we note the possibility of a quiescent $\log(M_{*}/M_{\odot}) < 8$ stellar mass galaxy along the quasar sightline below our continuum and flux density detection limits.
For component 2, we favour the scenario that it traces gas part of galaxy B's disc due to its alignment in velocity with the redshifted side of the galaxy.

Recent studies from MEGAFLOW \citep{Schroetter2016} and MUSE-ALMA Halos \citep{Hamanowicz2020} have also found multiple galaxies associated with \ion{Mg}{ii} and Lyman-$\alpha$ absorbers, respectively.
The system studied here is found to be similar in redshift, spatial and SFR distribution to galaxies associated with Lyman-$\alpha$ absorbers \citep{Hamanowicz2020}.
However, while \ion{Mg}{ii} and Lyman-$\alpha$ traces cool $1000$ K gas, the $21$-cm absorption line is more sensitive to cold gas that is likely to collapse into $\text{H}_2$, and subsequently form stars.
Typically, intervening $21$-cm \ion{H}{i} absorbers are associated with the inner disc of galaxies due to declining detection rates at higher impact parameters \citep{Borthakur2016,Curran2016,Dutta2017}. 
In the case of PKS 1610-771, it is clear from the opposite signs of the projected ionised and measured neutral gas velocities in front of the QSO that this is not the case.
If components 1a and 1b are indeed extragalactic gas clouds, this system emphasises the impact of interactions on how galaxies evolve as cold gas required for star formation is being removed. 
The direct connection between cold \ion{H}{i} and star formation allows us to draw more tangible links between gas processes such as outflows, inflows and stripping, and the effects of these processes on how galaxies evolve.

\section{Conclusion}
Through the novel and powerful combination of ASKAP and MUSE instruments, the nature of gas in this system has been successfully probed.
The neutral hydrogen absorption detection with the Australian Square Kilometre Array Pathfinder telescope reveals two components (1a and 1b) near $z = 0.4503$ separated by $30\ \text{km s}^{-1}$ \citep{Sadler2020}.
From the MUSE spectrum of QSO PKS 1610-771, strong absorption from ions \ion{Ca}{ii} and \ion{Na}{i} are found.
A component of these low-ionisation metals coincides with the neutral gas detected by ASKAP, while the other component (2) reveals another cloud of gas redshifted by $\sim 110\ \text{km s}^{-1}$ from the first.
Component 2 is likely of lower column density or higher temperature, and thus, is not found in the initial radio detection.
In total, three components of gas are found in front of PKS 1610-771 due to an overlap between the neutral hydrogen and low-ionisation metal components.

Imaging of the system from \citet{Courbin1997} reveals four galaxy-like objects (A, B, C and D) near PKS 1610-771.
From the MUSE observations, we find object C to be a faint M-type star.
To uncover the redshift of galaxy D outshone by the bright QSO, a 1D spectral PSF subtraction is performed and reveals its redshift to likely be $z \sim 0.5$ using corroborating evidence from \citet{Hunstead1980}.
While objects C and D are not associated with the absorber, we find two other galaxies in the field (X and Y) located at a projected distance of more than $100$ kpc from the quasar.
In total, there are four objects found at the redshift of the DLA.

Galaxies A and B are located at projected distances of $8.82$ and $17.8$ kpc from the absorber respectively, while X and Y are further afield ($136$ and $190$ kpc respectively).
In velocity space, galaxies X and Y have separations of $200$ and $520\ \text{km s}^{-1}$, respectively, from $z_{\rm abs} = 0.4503$.
In contrast, galaxies A and B are separated by $60$ and $20\ \text{km s}^{-1}$ from the DLA redshift.
Curiously, three of the four galaxy spectra (A, B and Y) have higher-order Balmer lines, suggesting an enhanced  period of star formation in the past several hundred Myrs caused by interactions.
From fitting the ionised gas kinematics, there is clear rotation in the ionised disc of galaxy B.
Galaxies A, X and Y have emission lines with low S/N or are too compact to fit for resolved kinematic maps.

The three components of gas in front of the quasar have velocities of $-40$ (1a), $-10$ (1b) and $+110 \ \text{km s}^{-1}$ (2) with respect to the systemic velocity of galaxy B.
Component 2 likely arises from co-rotation with galaxy B's disc as the redshift side of the ionised gas aligns with the QSO sightline.
In contrast, components 1a and 1b have the opposite sign to the projected rotational velocity for galaxy B and must originate from other processes. 
We find that extragalactic gas clouds are the most likely explanation for these two components, with galaxies A and B only separated by $\sim 17$ kpc and showing signs of intense star formation in the past several hundred Myrs.
Outflowing gas from galaxy B ($\Phi = 57\degree$) may also be responsible for the \ion{H}{i} absorber, but it remains unclear whether cold, dense neutral gas can survive entrained in hot wind material $20$ kpc from the galaxy centre without breaking down.

The gas traced in absorption is ultimately a combination of the scenarios mentioned above.
While exact identification of the phenomena responsible for the absorption features seen in the ASKAP and MUSE spectra are uncertain, this work already illustrates the intricacies of cold gas behaviour at a largely unexplored redshift.
With the emergence of large \ion{H}{i} surveys such as the First Large Absorption Survey in \ion{H}{i} \citep{Allison2020,allison2021} and the MeerKAT Large Absorption Line Survey (MALS) \citep{Gupta2016}, the amount and kinematics of cold neutral gas in hundreds of systems will be determined at redshift $z > 0.5$.
Contained within each of these individual detections is a puzzle waiting to be unravelled: what are the origins of the neutral gas in relation to its associated galaxies?
If answered within enough systems by combining radio data with optical or millimetre observations, we are able to gain insights into the impact of gas on galaxy evolution during an era in the Universe not well understood.

\section*{Data Availability}
The data underlying this article will be shared on reasonable request to the corresponding author.

\section*{Acknowledgements}
We thank Jianhang Chen for checking for dust-continuum detections in ALMA calibrator data. We thank the anonymous referee for their comments.

We acknowledge the financial support of the Australian Research
Council through grant CE170100013 (ASTRO3D).
This research is supported by an Australian Government Research Training Program (RTP) Scholarship.
RZS acknowledges support from the National Science Foundation of China (11873073).
Based on observations collected at the European Southern Observatory under ESO programme 0103.A-0656.
We thank Thomas Ott for developing and distributing the \textsc{QFitsView} software.
This research made use of Astropy,\footnote{http://www.astropy.org} a community-developed core Python package for Astronomy \citep{astropy:2013, astropy:2018}.

%%%%%%%%%%%%%%%%%%%%%%%%%%%%%%%%%%%%%%%%%%%%%%%%%%



\bibliographystyle{mnras}
\bibliography{FLASH_MUSE_1610}
%%%%%%%%%%%%%%%%%%%%%%%%%%%%%%%%%%%%%%%%%%%%%%%%%%

%%%%%%%%%%%%%%%%% APPENDICES %%%%%%%%%%%%%%%%%%%%%

\appendix

\section{Synthetic continuum-subtracted narrowband}
\begin{figure}
    \includegraphics[width=\columnwidth]{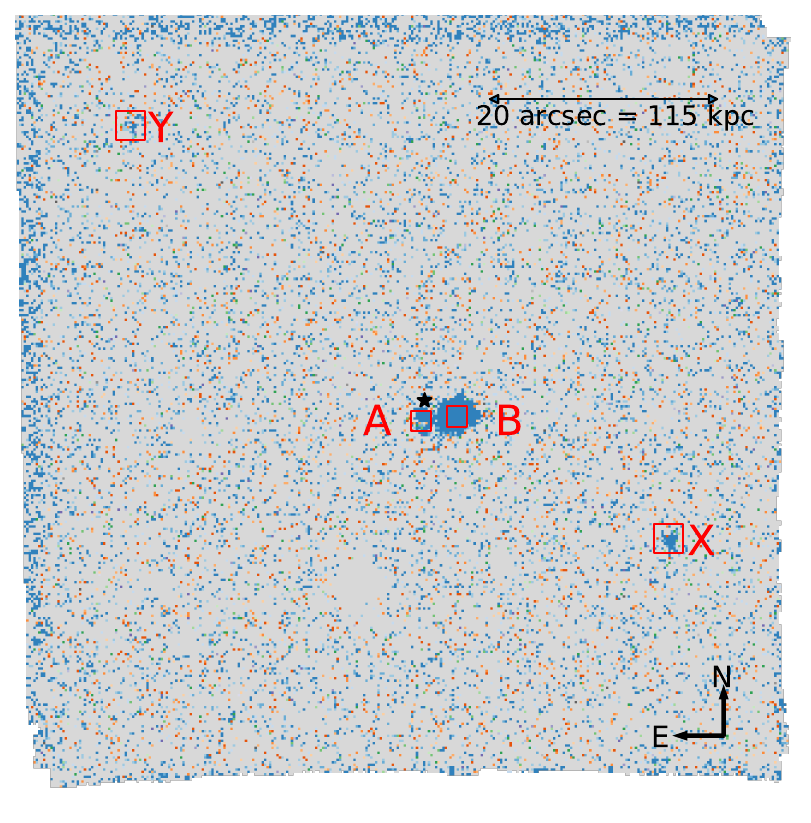}
    \caption{Synthetic continuum-subtracted narrow-band image with width $1000$ km s$^{-1}$ around [\ion{O}{ii}] doublet at $z_{\rm abs}=0.4503$.
    The background continuum is estimated by applying a median filter on a region with width $3000$ km s$^{-1}$ offset from the edges of the narrow-band by $10$ pixels.
    Orientation is north-east (up-left).
    Galaxies associated with the \ion{H}{i} $21$-cm absorber have been marked in red, and the centre of quasar PKS 1610-771 by a black star.
    In addition to objects A and B, we find two galaxies (labelled X and Y) with projected distances more than $150$ kpc from the absorber.
    {The position of the group members is a possible filamentary structure stretching from galaxy Y (north-east) to galaxy X (south-west).}}
    \label{fig:OII_NB}
\end{figure}

Fig. \ref{fig:OII_NB} depicts a synthetic continuum-subtracted narrow-band (NB) image centred around the [\ion{O}{ii}] doublet at the redshift of the absorber $z_{\rm abs} = 0.4503$.
It reveals four objects with [\ion{O}{ii}] emission near $z_{\rm abs} = 0.4503$ and these are marked in red (A, B, X and Y).
In addition to galaxies A and B, which were known to be associated with the DLA from optical spectroscopy \citep{Sadler2020}, we find another two galaxies (X and Y) with impact parameters of $136$ and $190$ kpc respectively from the QSO sightline.
This galaxy overdensity is aligned in what may be a filamentary structure extending from north-east to south-west.

\section{Objects C and D}
\label{sec: appC}
\begin{figure}
  \centering
      \includegraphics[width=\columnwidth]{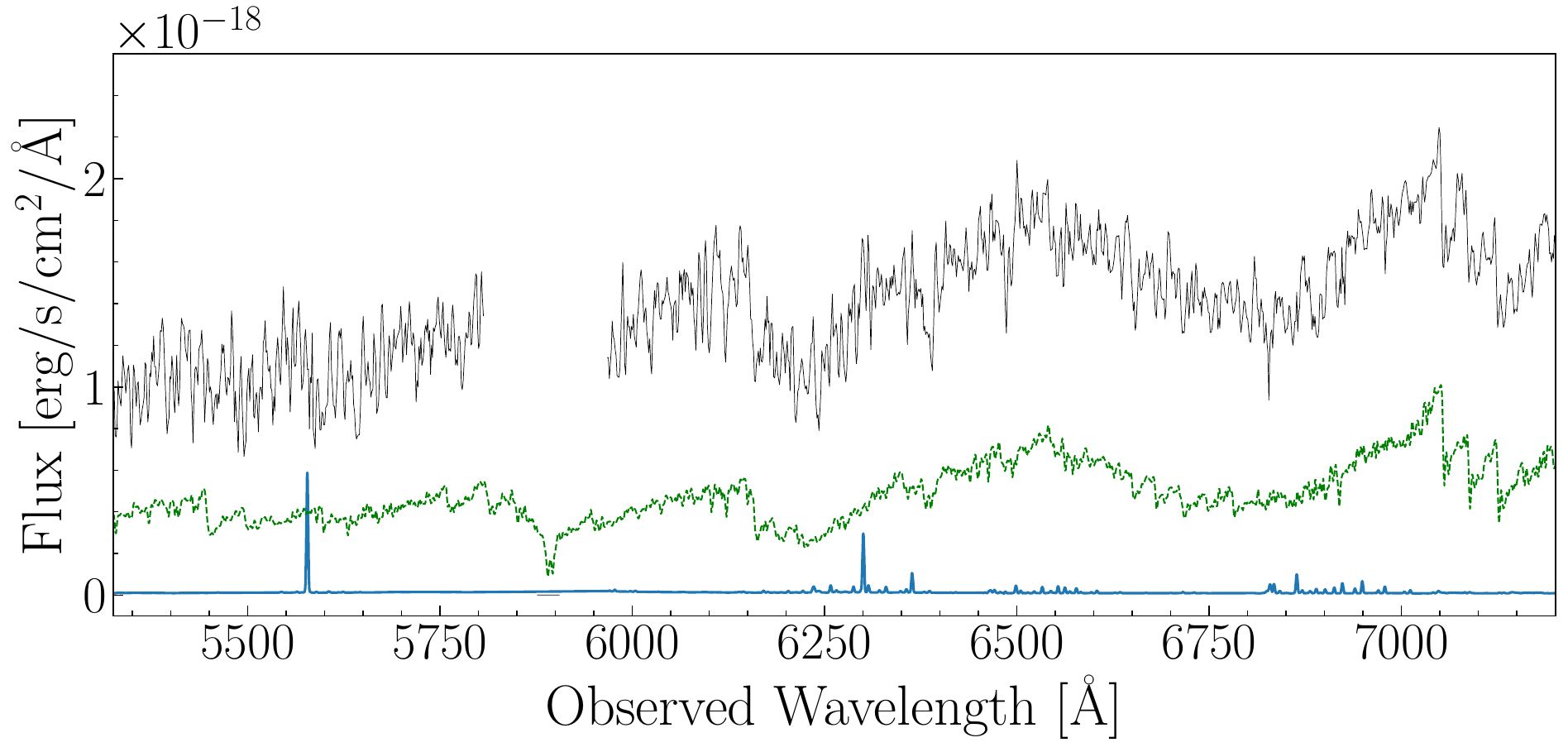}
  \caption{A smoothed ($3$ pixel moving average) spectrum of object C is plotted in black with the corresponding sky spectrum in blue (not to scale).
   An arbitrarily scaled spectrum of M5V type star HD 173740 is plotted in green for comparison.}
  \label{fig:galC}
\end{figure}

Despite the proximity of objects C and D to the absorber on the sky, neither is associated with the DLA.
The former is evidently a faint M-type star from Fig. \ref{fig:galC}.

\begin{figure}
  \centering
  {\includegraphics[width=0.48\columnwidth]
  {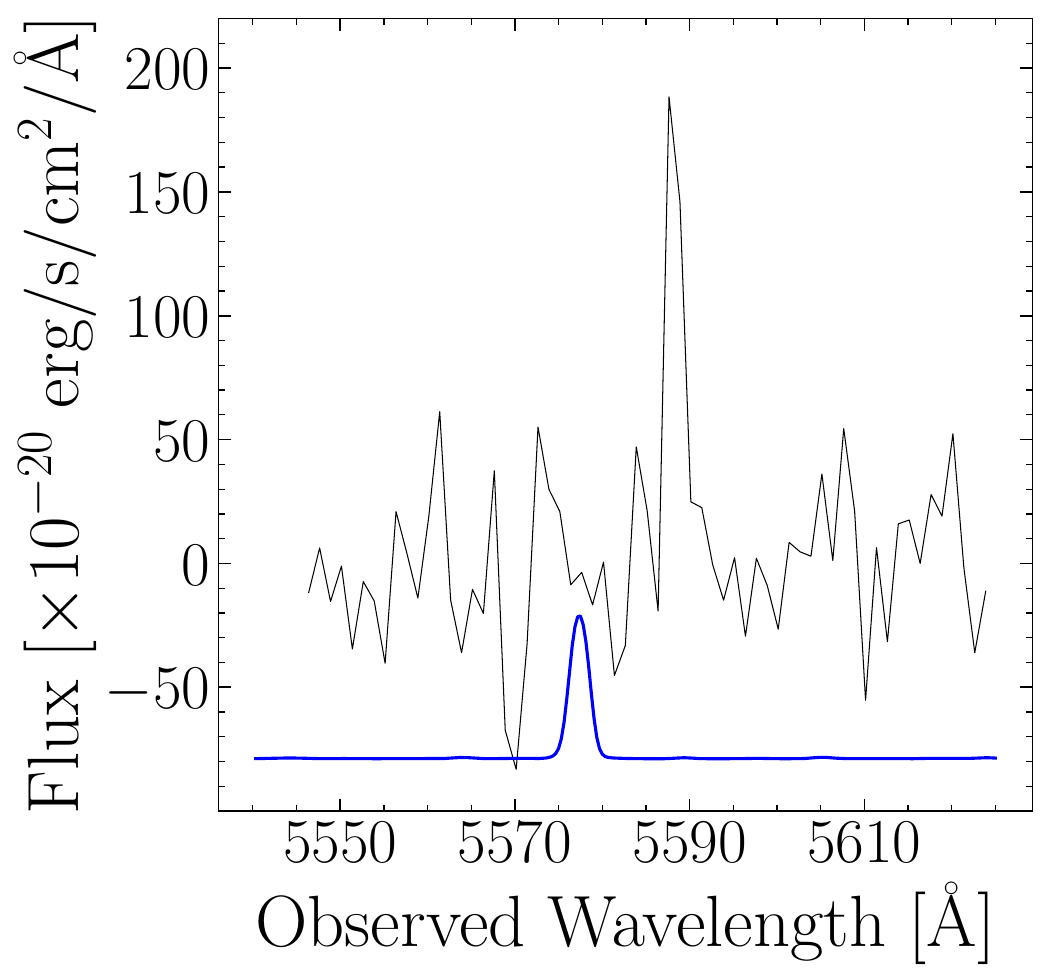}}
  \hfill
  {\includegraphics[width=0.48\columnwidth]
  {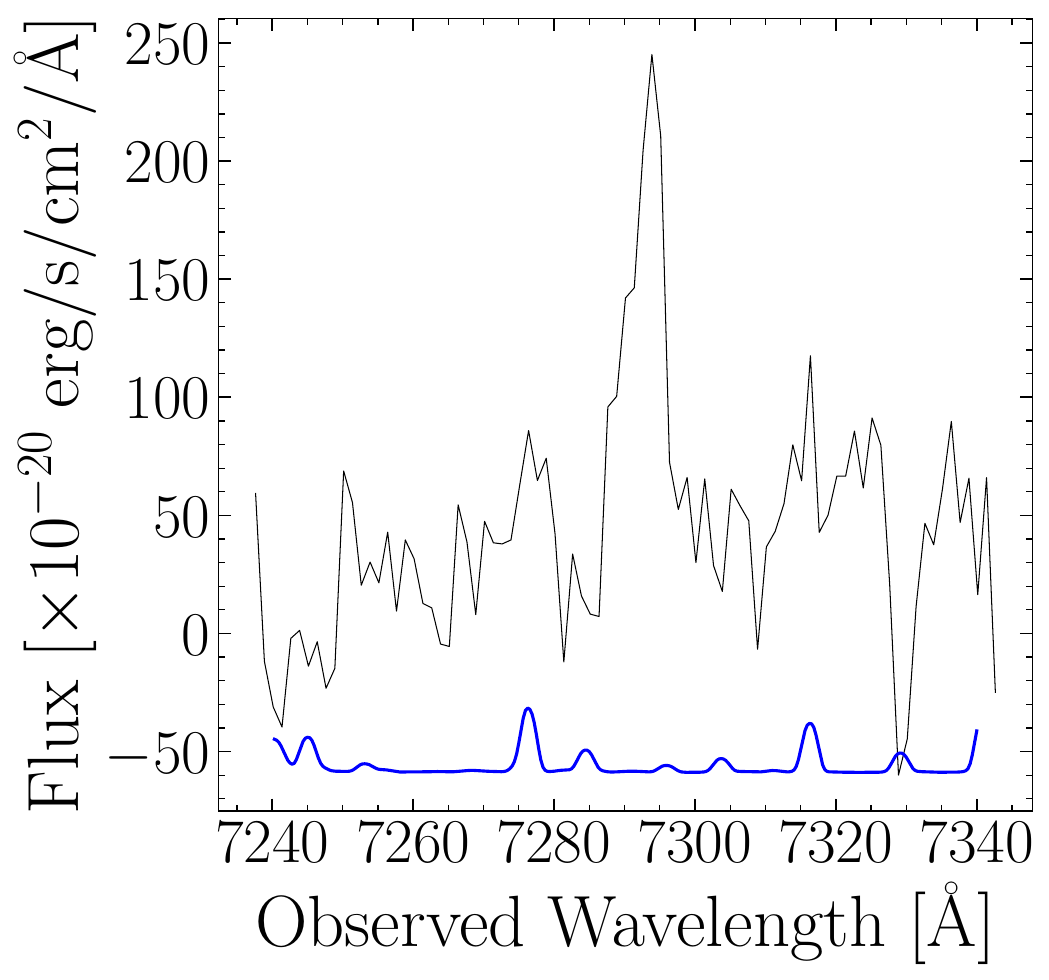}}
  \caption{Candidate emission lines \ion{O}{ii} (left) and H$\beta$ (right) found in front of QSO PKS 1610-771 after a 3D PSF subtraction (black).
  Scaled and vertically offset sky spectrum is in blue.
  The best-fitting redshift is $z = 0.5001$ for the pair of lines.}
  \label{fig:Gal_D_lines}
\end{figure}

After the 3D PSF subtraction to unearth object D, there are two candidate emission lines in its spectrum near $5587$ and $7290$ \AA, and these are depicted in Fig. \ref{fig:Gal_D_lines}. 
The two emission lines are consistent with [\ion{O}{ii}] and H$\beta$ at $z = 0.5001$.
Thus, galaxy D is unlikely to be associated with the DLA.

\section{Galaxy Spectra}
\label{sec: appB}
The MUSE spectra of galaxies A, B, X and Y are displayed in this appendix.
For three of the objects (A, B and Y), the stellar continuum is clearly detected, while emission lines are prevalent in all spectra.
Poorer sky subtraction near the edges of the MUSE field result in artefacts at $6500$ and $6850$ \AA \ in the spectra of galaxies X and Y.
The spectral gap from $5820$ \AA \ to 5970 \AA \ correspond to a notch filter between 5820-5970 Å inherent to AO-assisted observations.

\begin{figure}
  \centering
      \includegraphics[width=\columnwidth]{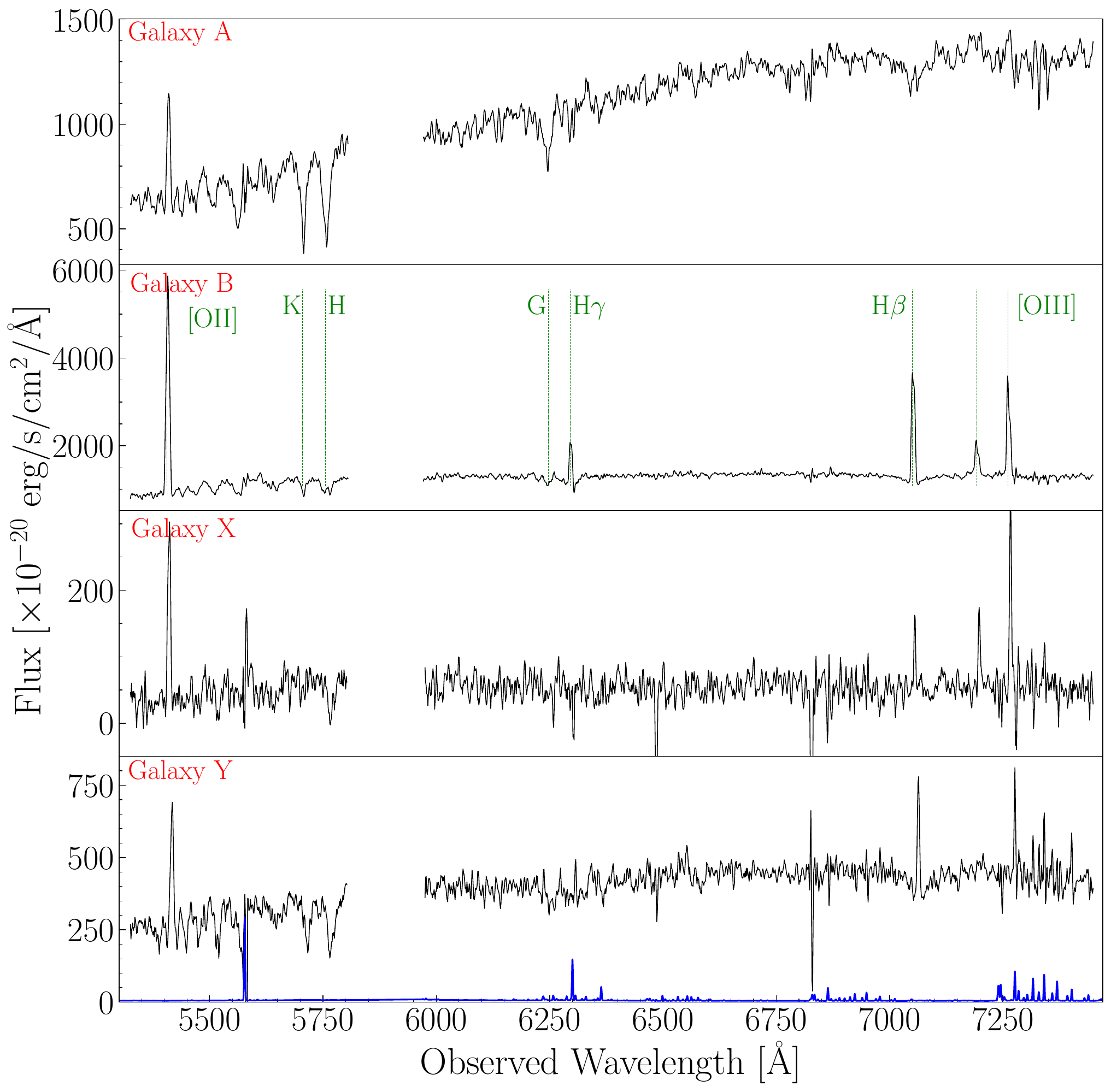}
  \caption{Spectra of the four galaxies associated with the $z = 0.45$ absorber (A, B, X and Y from top to bottom) from MUSE.
  Emission and absorption lines are marked on the spectrum of galaxy B and can also be seen in the remaining group members.
  A scaled and vertically offset sky spectrum is included in blue.
  Each spectrum has been smoothed using a 3 pixel moving average.}
  \label{fig:all_gal_spec}
\end{figure}

\section{PKS 1610-771 Metal Lines}
\begin{figure*}
  \centering
      \includegraphics[width=0.9\textwidth]{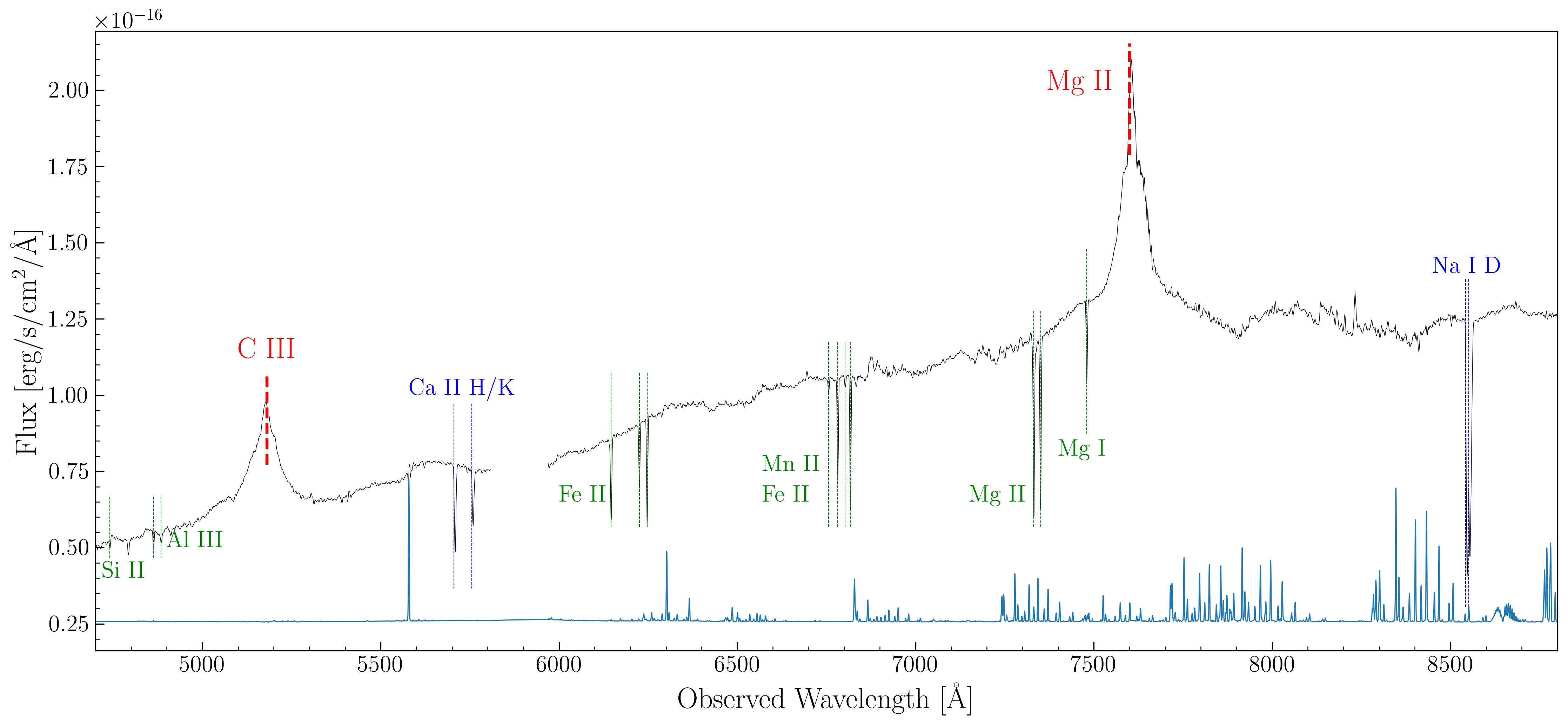}
  \caption{A smoothed ($3$ pixel moving average) spectrum of the QSO with the corresponding sky spectrum in blue (not to scale).
   Marked in red are the broad emission lines of the quasar (\ion{Mg}{ii} and \ion{C}{iii}) at $z = 1.71$.
   A separate metal system (green) is found at redshift $z=1.621$, but this is not associated with the DLA.
   Instead, the \ion{Ca}{ii} and \ion{Na}{i} metal absorption lines marked in blue are at the redshift of the $z = 0.45$ absorber.}
  \label{fig:QSO_spec}
\end{figure*}

Another system at redshift $z = 1.621$ can be seen in the the background quasar through various absorption lines (\ion{Si}{II}, \ion{Al}{iii}, \ion{Fe}{ii}. \ion{Mn}{ii}, \ion{Mg}{ii} and \ion{Mg}{i}). 
These are marked in green in Fig. \ref{fig:QSO_spec} along with the broad emission lines of the QSO (red), and \ion{Ca}{ii} and \ion{Na}{i} absorption linked to our \ion{H}{i} detection.
However, the significant disparity in redshift means this higher redshift system is not related to our absorber.

%%%%%%%%%%%%%%%%%%%%%%%%%%%%%%%%%%%%%%%%%%%%%%%%%%

% Don't change these lines
\bsp	% typesetting comment
\label{lastpage}
\end{document}